\documentclass[a4paper, titlepage, 10pt, usenames, dvipsnames, svgnames, table, final]{article}

\usepackage{geometry}

\usepackage{xr}
\usepackage{graphicx}
\usepackage[utf8]{inputenc}

\usepackage{color}

\usepackage{siunitx}
\DeclareSIUnit\kgperL{\kilo\gram\per\liter}
\DeclareSIUnit\waveno{\per\centi\metre}
\DeclareSIUnit\invcm{\per\centi\metre}
\DeclareSIUnit\diffconst{\square\centi\metre\per\second}
\DeclareSIUnit\VpA{\volt\per\angstrom}

\usepackage[version=3]{mhchem}

\usepackage{caption}
\usepackage{subcaption}

\usepackage{multicol}

\usepackage{verbatim}

\usepackage{revsymb}

\newcommand{\fref}[1]{Fig.~\ref{#1}}
\newcommand{\sref}[1]{Sec.~\ref{#1}}
\newcommand{\eref}[1]{Eq.~\ref{#1}}

\usepackage[hyperref=false,
            url=false,
            isbn=false,
            doi=false,
            eprint=false,
            sorting=none,
            sortcites=true,
            firstinits=true,        % Print all authors name in initials (first and middle names)
            maxbibnames=99,
            backend=biber]{biblatex}
\AtEveryBibitem
{
    \clearfield{number}
    \clearfield{issue}
    \clearfield{eid}
    \clearfield{note}
}
\DefineBibliographyExtras{english}{}
\DefineBibliographyExtras{english}{}
\bibliography{literature}
\renewbibmacro{in:}{}

\usepackage{authblk}

\makeatletter
\renewcommand*{\@fnsymbol}[1]{\ifcase#1\or*\else\@arabic{\numexpr#1-1\relax}\fi}
\makeatother

\renewcommand{\vec}[1]{\mathbf{#1}}

\newcommand{\ramanint}{e^4a_0^4 E_h^{-2} \si{\centi\metre}}

\begin{document}

\title{Mimyria: Machine learned vibrational spectroscopy for aqueous systems made simple}

\author[1,2]{Philipp Schienbein\thanks{email: philipp.schienbein@ruhr-uni-bochum.de}}
\affil[1]{Lehrstuhl für Theoretische Chemie II, Ruhr-Universität Bochum, 44780 Bochum, Germany}
\affil[2]{Research Center Chemical Sciences and Sustainability, Research Alliance Ruhr, 44780 Bochum, Germany}
\date{ }

\maketitle

\begin{abstract}
    Vibrational spectroscopy provides a powerful connection between molecular dynamics (MD) simulations and experiment, but its routine use in condensed-phase systems remains limited.
    We introduce \texttt{mimyria}, a modular and automated framework that orchestrates electronic-structure reference calculations,
    trains atom-resolved machine-learning response models, and generates IR and Raman spectra from MD trajectories within a unified workflow.
    We introduce the polarizability gradient tensor (PGT) as a novel atom-resolved machine-learning target property for Raman spectroscopy, complementing the established atomic polar tensor (APT) for IR spectroscopy. 
    As a necessary prerequisite, we demonstrate how both PGTs and APTs can accurately be computed from electronic-structure theory, validate them across formally equivalent derivative formulations, and thereby benchmark their numerical consistency.
    We then employ machine learning as an efficient surrogate to represent the validated APT and PGT response functions on aqueous benchmark systems. 
    We validate the trained models directly at the level of the vibrational spectrum against explicit \emph{ab initio} reference calculations and find that IR and Raman spectra converge with surprisingly small training sets. 
    Moreover, spectral agreement improves more rapidly than the root-mean-square error (RMSE), the conventional model error metric. 
    While RMSE is straightforward to compute, statistically converged reference spectra are generally impractical to obtain, motivating the need to relate model-level errors to observable-level accuracy.
    By connecting these complementary error measures, we provide practical guidelines and early-stopping criteria for achieving sufficient spectral fidelity. 
    By integrating response-tensor learning, automated training, and spectral-domain validation into a unified workflow, \texttt{mimyria} enables data-efficient and quantitatively reliable vibrational spectroscopy.
\end{abstract}

\newcommand{\field}{\varepsilon}

\section{Introduction}

Vibrational spectroscopy provides a direct connection between theory
and experiment, as it 
represents a physical observable that 
is accessible in both domains~\cite{Bakker-2010-ChemRev, Perakis-2016-ChemRev}.
Derived from  molecular dynamics~(MD) simulations, vibrational spectra explicitly include thermal and anharmonic effects, as well as the influence of thermal fluctuations in condensed-phase systems.
As a result, qualitative~-- and in some cases even quantitative~\cite{Mauelshagen-2025-SciAdv}~-- agreement with experimental data can be achieved.
Vibrational spectra are therefore used to validate performed MD simulations because it probes both the structure and the dynamics of the simulated system.
Beyond its use as a powerful validation tool, condensed-phase theoretical vibrational spectroscopy yields rich physical information: 
spectral features can be associated with specific atomic motions~\cite{Heyden-2010-PNAS, Galimberti-2019-FaradayDiscuss, Vuilleumier-2023-CondensMatterPhys, Flor-2024-Science, Joll-2025-JPCL-THz}, 
and molecules in distinct chemical environments,
such as interfaces or solvation shells~\cite{Perera-2009-PNAS, Heisler-2010-Science, Perakis-2016-ChemRev, Schienbein-2017-JPCL, BenAmotz-2019-JACS, Schwaab-2019-ANIE},
can be distinguished. 
In addition, dynamical time scales, including reorientation times and collision rates, can be extracted and related to experimental observations~\cite{Mauelshagen-2025-SciAdv}.
Access to such atom-resolved and environment-specific insights relies on the ability to decompose vibrational spectra into contributions from individual atoms or molecules~\cite{Gaigeot-2003-JPCB, Martinez-2006-JCP, Heyden-2010-PNAS, Sun-2014-JACS, Schienbein-2017-JPCL, Imoto-2019-JCP, Galimberti-2019-FaradayDiscuss, Joll-2025-JPCL-THz}.
Within IR spectroscopy, 
one powerful route is 
through the atomic polar tensor (APT)~\cite{Person-1974-JCP}, also called ``Born effective charge tensor''~(BEC)~\cite{Gonze-1992-PRL}, which enables the total spectrum to be rigorously expressed in terms of atomic spectral responses~\cite{Gaigeot-2007-MolPhys, Khatib-2017-JPCL, Imoto-2019-JCP, Schienbein-2023-JCTC, Joll-2025-JPCL-THz},
without relying on charge partitioning schemes.
This decomposition allows, for instance, vibrational spectra to be analyzed in terms of translational, rotational, and intermolecular vibrational contributions~\cite{Joll-2025-JPCL-THz}, thereby providing a direct link between atomic dynamics and spectral signatures.

Condensed-phase theoretical vibrational spectroscopy, however, is not yet routinely employed to extract physical insights. 
A primary reason is the substantial computational effort required to obtain statistically converged spectra, which often demands hundreds of picoseconds of MD trajectories, particularly 
when targeting low-intensity spectral features or difference spectra~\cite{Schmidt-JACS-2009, Kann-2016-JCP, Schienbein-2017-JPCL}.
The calculation of vibrational spectra from \emph{ab initio} MD was pioneered long ago~\cite{Silvestrelli-1997-ChemPhysLett}, 
but 
such long simulations remain computationally demanding.
For atom-resolved spectral analysis,
additional electronic structure evaluations are required when spectra are decomposed into atomic contributions using APTs, because
these tensors
must be evaluated sufficiently frequently along the trajectory, ideally at every time step~\cite{Galimberti-2017-JCTC}.
Various more or less severe approximations have therefore been explored, ranging from parametrized APT models and instantaneous-mode approximations to evaluating APTs only intermittently along the trajectory~\cite{Galimberti-2017-JCTC, Khatib-2017-JPCL, Imoto-2019-JCP}.
For Raman spectroscopy, this challenge is further exacerbated by the need to compute polarizability tensors, which are typically obtained via perturbation theory or numerical derivatives and therefore constitute an additional computational cost. 
In \emph{ab initio} MD simulations, polarizability tensors are therefore often computed only partially, for instance along a single Cartesian component~\cite{Thomas-2013-PCCP}. 
As a result, rigorous atom-resolved decompositions of Raman spectra remain rare in the current literature.

Condensed-phase theoretical vibrational spectroscopy, however, is not yet routinely employed to extract physical insights.
A primary reason is the substantial computational effort required to obtain statistically converged spectra, which 
often demands hundreds of picoseconds of MD trajectories, 
particularly when the spectral responses of specific chemical species must be isolated from a dominant background~\cite{Kann-2016-JCP, Schienbein-2017-JPCL}.
While calculating vibrational spectra from \emph{ab initio} MD 
was pioneered long ago~\cite{Silvestrelli-1997-ChemPhysLett},
such simulations require a significant computational effort.
For Raman spectroscopy, this challenge is further exacerbated by the need to compute polarizability tensors, which are typically obtained via 
perturbation theory or numerical derivatives and therefore constitute an additional computational cost. 
In \emph{ab initio} MD simulations, polarizability tensors are therefore often computed only partially, for instance along a single Cartesian component~\cite{Thomas-2013-PCCP}.
As a result, rigorous atom-resolved decompositions of Raman spectra remain rare in the current literature.

With the aim of accelerating \emph{ab initio} MD by several orders of magnitude while retaining the \emph{ab initio} quality, machine learning~(ML) potentials
have been introduced that replace expensive electronic-structure calculations with computationally efficient models~%
\cite{Behler-2007-PRL, Behler-2021-ChemRev, Bartok-2010-PRL, Batzner-2022-NatCommun, Batatia-2022-NeurIPS, Bochkarev-2024-PRX}. 
Most commonly used ML potentials, however, do not automatically provide the electronic response functions required for the calculation of vibrational spectra. 
As a consequence, additional ML models must be trained or augmented and retrained to represent these response properties before IR or Raman spectra can be generated.

These challenges demand approaches that provide access to vibrational response functions at atomic resolution without requiring long \emph{ab initio} trajectories. 
Motivated by the dissective power of the APT, we previously introduced a ML model that directly represents atom-resolved APTs, referred to as the atomic polar tensor neural network (APTNN)~\cite{Schienbein-2023-JCTC}.
At the time, most ML approaches for vibrational spectroscopy focused on learning global response properties, such as total dipole moments or polarizability tensors~%
\cite{Gastegger-2017-ChemSci, Sifain-2018-JPCL, Wilkins-2019-PNAS, Raimbault-2019-NewJPhys, Gastegger-2021-ChemSci, Kapil-2024-Faraday, Xu-2024-JCTC},
which are physically defined for the system as a whole rather than for individual atoms. 
Several works have now adopted the capability to represent APTs as well~\cite{Kapil-2024-Faraday, Schmiedmayer-2024-JCP, Stocco-2025-npj}; 
however, in most cases these are obtained indirectly as derivatives of a learned total dipole moment, 
in close analogy to ML potentials where forces are usually derived as gradients of the potential energy.
In contrast, the central idea of the APTNN is to learn atomic APTs directly, thereby avoiding the need to train a total dipole moment and consequently avoiding the non-unique decomposition of that global object into atomic contributions.
The direct-derivative-learning strategy is therefore largely complementary to approaches that focus on global response properties.
Both training strategies have been demonstrated to achieve comparable accuracy, but they emphasize on different aspects of the underlying physics.
Direct derivative learning exploits that the gradients are physically and gauge- and branch-invariant response quantities and therefore do not suffer from the multi-valuedness of, for instance, the dipole moment in periodic systems~\cite{Spaldin-2012-JSolidStateChem}. 
Moreover, we recently demonstrated that 
accurate IR spectra of bulk liquid water 
can be produced by
using training data obtained exclusively from finite gas-phase water clusters~\cite{Jindal-2025-JCTC}. 
In this setting, a total dipole moment cannot be meaningfully transferred between finite and periodic systems, whereas the APT, as a size-insensitive property, can 
be converged for the central atoms in a sufficiently large finite cluster and transferred to the periodic bulk environment.
We note in passing that APTs have also recently been explored in the context of incorporating long-range electrostatics into ML potentials~%
\cite{Zhong-2025-npj, Staerk-2026-arxiv}. 
Furthermore, APTs were used to incorporate external electric fields in MLMD simulations~\cite{Joll-2024-NatCommun, Stocco-2025-npj}.
In the present work, we further show that closely related ideas can be extended to Raman spectroscopy, introducing the so-called ``polarizability gradient tensor''~(PGT) as ML target.

Another important aspect concerns the validation of vibrational spectra generated using ML models. 
Obtaining statistically converged vibrational spectra typically requires several tens to hundreds of picoseconds of MD trajectories. 
When explicit \emph{ab initio} reference calculations are employed, such trajectory lengths constitute a substantial computational commitment, and even spectra with deliberately reduced statistical accuracy still rely on simulation times that are costly 
when using explicit electronic-structure calculations.
As a consequence, one has to accept that statistically converged \emph{ab initio} reference spectra are generally unavailable in practice, and that even reference spectra with minimal statistical accuracy are difficult to obtain, in particular for large systems or when computationally demanding electronic-structure methods are required. 
This limitation becomes especially severe for system sizes beyond a few hundred to thousand atoms, where explicit \emph{ab initio} calculations are effectively impractical.
The central question is therefore whether the accuracy of vibrational spectra can be inferred from the ML model itself, without computing statistically converged \emph{ab initio} reference spectra.
Several strategies addressing related questions have been proposed in the literature~\cite{Schran-2020-JCP, Schran-2021-PNAS, Stolte-2025-JCTC}, including active-learning approaches, but primarily in the context of assessing ML potential rather than electronic response properties.

Herein, we present a complete workflow that connects MD trajectories to vibrational spectra. 
The proposed software framework (``\texttt{mimyria}'') provides the necessary tools to train ML models for electronic response functions and to post-process MD trajectories to obtain IR and Raman spectra.
The generation of training data, the training of response models, and the subsequent calculation of vibrational spectra are handled within a unified and largely automated workflow, requiring only minimal user intervention. 
The ML models are intentionally designed in a modular fashion, such that they do not interfere with the ML potentials used to generate the MD trajectories. 
As a result, models for IR and Raman spectra can be trained and applied independently. 
This modularity offers significant practical flexibility, as numerous ML potential have been trained on various different systems in recent years and can now be revisited to generate vibrational spectra without retraining the underlying interaction model. 
For future projects it is, moreover, not necessary to decide at the outset of a project whether vibrational spectra will be required; the corresponding response models can be trained at a later stage when such analyses become relevant.
We note in passing that electronic response functions can, in principle, also be obtained as higher-order derivatives of the potential energy~\cite{Gastegger-2021-ChemSci}. 
While this represents an elegant approach, it requires the underlying potential to be revisited and revalidated once vibrational spectra become relevant.

We demonstrate the applicability of 
\texttt{mimyria}
by computing IR and Raman spectra of aqueous systems. 
In particular, we exemplarily build on existing ML potential for liquid water~\cite{Joll-2024-NatCommun} and an aqueous sulfate solution~\cite{Schran-2021-PNAS}
to illustrate how 
accurate training data can be obtained from electronic structure calculations,
APTs and PGTs 
can be trained, and vibrational spectra can be calculated within \texttt{mimyria}.
Finally, we illustrate how the quality of atom-resolved spectral predictions can be assessed for rarely occurring atomic environments, such as 
the sulfate ion in this case, 
that can be dominated by the background in the total spectrum.
These examples highlight the modular and complementary nature of the framework, as the underlying interaction potential can be employed without retraining or modification. 
Finally, we address how the spectroscopic error introduced by the ML models can be quantified, thereby closing the loop from MD trajectories to validated vibrational spectra.

\section{Methods}

\subsection{Calculating IR Spectra}

The calculation of IR absorption coefficients has been detailed many times~\cite{McQuarrie2000, Heyden-2010-PNAS, Thomas-2013-PCCP}, 
also in the context of APTs~\cite{Schienbein-2023-JCTC}, but is briefly summarized herein to introduce the nomenclature.
The frequency dependent Beer-Lambert absorption coefficient of IR spectroscopy
\begin{equation}
    \alpha(\omega) = 
    \frac{\pi\beta\omega^2}{3 V c \epsilon_0 n(\omega)} \frac{1}{2\pi} 
    \int_{-\infty}^\infty \, dt \, e^{-i\omega t}
    \left<
    \vec{M}(0) \vec{M}(t)
    \right>
\end{equation}
is calculated using the time auto correlation function of the total dipole moment $\vec{M}(t)$, 
where $\beta = 1/k_\text{B}T$,
$k_\text{B}$ is the Boltzmann constant, 
$T$ the temperature, 
$V$ the volume of the simulation box, 
$c$ the speed of light in vacuum,
and
$n(\omega)$ the frequency dependent refractive index.
The Kubo transform of the quantum time correlation function is already included in the equation that is sometimes also called ``harmonic quantum correction factor''~\cite{Kubo-1991-StatPhysII, Ramirez-2004-JCP, Pabst-2025-arxiv}.
Using an exact identity of Fourier transforms and time derivatives, the equation can be exactly rewritten as 
\begin{equation}
    \alpha(\omega) = 
    \frac{\pi\beta}{3 V c \epsilon_0 n(\omega)} \frac{1}{2\pi} 
    \int_{-\infty}^\infty \, dt \, e^{-i\omega t}
    \left<
    \dot{\vec{M}}(0) \dot{\vec{M}}(t)
    \right>
    \, ,
    \label{eq:ir_from_Mdot}
\end{equation}
now using the time derivative of the total dipole moment $\dot{\vec{M}}(t)$.
Using the chain rule, the total dipole moment time derivative can exactly be rewritten~\cite{Schienbein-2023-JCTC}
\begin{equation}
    \dot{M}_\zeta (t)
    = 
    \sum_{i,\eta} 
    \left.\frac{\partial M_\zeta}{\partial r_{i,\eta}}\right|_t 
    \left.\frac{\partial r_{i,\eta}}{\partial t}\right|_t
    = 
    \sum_{i,\eta}
    \mathcal{P}_{i, \eta, \zeta}(t) \cdot v_{i, \eta}(t)
    \, ,
    \label{eq:Mdot_from_apt}
\end{equation}
in terms of a sum running over all atoms $i$ in the system, 
where each APT,
$\vec{\mathcal{P}}_i$, 
is multiplied by the corresponding atomic velocity, 
$\vec{v}_i$.
The APT is a 3x3 tensor defined for each atom $i$.
The
indices $\eta$ and $\zeta$ represent Cartesian components of the atomic displacement and of an externally applied electric field, respectively~-- 
the latter will become apparent in \sref{sec:electronic_structure}.
Keeping the two kinds of indices strictly separated is helpful, because it makes all tensor multiplications visually transparent and removes any ambiguity about the dimensions involved in them.
Inserting \eref{eq:Mdot_from_apt} into \eref{eq:ir_from_Mdot} allows one to represent the absorption coefficient $\alpha(\omega)$ in terms of APTs and atom velocities.

\subsection{Calculating Raman Spectra}
\label{sec:methods-raman}
Similar to IR spectroscopy,
Raman spectra have been computed in the literature and the equations are again briefly summarized herein.
Raman scattering is a second order process that depends on the incident and scattered light beams,
particularly how the incident light is polarized and what polarization of the scattered light is collected. 
The non-resonant Raman differential scattering cross section into a frequency range $d\omega$ and a solid angle $d\Omega$ is given by~\cite{Gordon-1965-JCP, Gordon-1968-AdvMagResonance, McQuarrie2000}
\begin{equation}
    \frac{d^2\sigma}{d\omega d\Omega} 
    = 
    \lambdabar^{-4}
    \frac{1}{2\pi} 
    \int_{-\infty}^\infty \, dt \, e^{-i\omega t}
    \left<
    \left(\vec{\epsilon}_\text{i} \cdot \vec{\alpha}(0) \cdot \vec{\epsilon}_\text{s} \right)
    \left(\vec{\epsilon}_\text{i} \cdot \vec{\alpha}(t) \cdot \vec{\epsilon}_\text{s} \right)
    \right>
    \, ,
\end{equation}
where 
is the reduced 
wavelength of the scattered non-resonant laser light,
$\omega$ denotes the Raman shift corresponding to molecular vibrational frequencies,
$\vec{\alpha}(t)$ is the polarizability tensor of the system at time $t$, 
and
$\vec{\epsilon}_\text{i}$ and $\vec{\epsilon}_\text{s}$ are unit vectors describing the polarization of the incident and scattered light, respectively.
That equation 
can be transformed using the same Fourier-Transform identity employed for the IR absorption coefficient to yield
\begin{equation}
    \frac{d^2\sigma}{d\omega d\Omega} 
    = 
    \lambdabar^{-4}
    \frac{1}{\omega^2}
    \frac{1}{2\pi} 
    \int_{-\infty}^\infty \, dt \, e^{-i\omega t}
    \left<
    \left(\vec{\epsilon}_\text{i} \cdot \dot{\vec{\alpha}}(0) \cdot \vec{\epsilon}_\text{s} \right)
    \left(\vec{\epsilon}_\text{i} \cdot \dot{\vec{\alpha}}(t) \cdot \vec{\epsilon}_\text{s} \right)
    \right>
    \, ,
    \label{eq:raman-1}
\end{equation}
that uses
the time derivative of the polarizability tensor $\dot{\vec{\alpha}}(t)$.

When it comes to the absolute scattering intensity, the situation is unfortunately less clear when 
compared to the IR absorption coefficient, 
since reported Raman spectra strongly depend on both the experimental or theoretical setup.
Ordinarily, Raman scattering spectra are normalized such that constant prefactors vanish; 
even the factor $\lambdabar^{-4}$
is sometimes removed~\cite{Gordon-1965-JCP}, which alters relative intensities of peaks in a weak but frequency-dependent manner. 
Spectra may further be Bose--Einstein corrected~\cite{Sommers-2020-PCCP}, adding an additional frequency dependent prefactor $1-\text{exp}\left(-\beta\hbar\omega\right)$.
In theoretical work, the Kubo transform of the quantum mechanical correlation function is often employed which introduces the so-called ``harmonic quantum correction factor''~\cite{Ramirez-2004-JCP} in the scattering cross section~\cite{Thomas-2013-PCCP, Thomas-2015-PCCP, Ditler-2022-WIREsComputMolSci}.
This factor partially cancels the Bose-Einstein correction.
An alternative formulation expresses the Raman lineshape in terms of the imaginary part of the susceptibility%
~\cite{McQuarrie2000, Pabst-2025-arxiv}. 
Remarkably, the prefactor $1-\exp(-\beta\hbar\omega)$ enters explicitly,
but is canceled by the Kubo transform~\cite{Pabst-2025-arxiv}.
This procedure rigorously yields the power spectrum of the fluctuations of $\alpha(t)$ and thus assigning a proper intensity to that power spectrum.
This susceptibility power spectrum is, however, not identical to the Raman scattering cross section and therefore the problem of quantitative comparability between different works remains.

From a theoretical point of view, all molecular vibrational information is completely determined by the $\omega$-dependent time-correlation function of the polarizability, 
whereas the $\lambdabar$-dependent optical prefactors and the overall normalization only affect how this information is experimentally detected and scaled.
It
is thus pragmatic to introduce a Raman scattering lineshape function 
\begin{equation}
    I'_\text{i,s} (\omega)
    \propto
    \frac{1}{\omega^2}
    \frac{1}{2\pi} 
    \int_{-\infty}^\infty \, dt \, e^{-i\omega t}
    \left<
    \left(\vec{\epsilon}_\text{i} \cdot \dot{\vec{\alpha}}(0) \cdot \vec{\epsilon}_\text{s} \right)
    \left(\vec{\epsilon}_\text{i} \cdot \dot{\vec{\alpha}}(t) \cdot \vec{\epsilon}_\text{s} \right)
    \right>
    \, ,
\end{equation}
that contains all relevant Raman signals, while discarding absolute normalization and optical prefactors. 
The latter can subsequently be reinstated as needed for comparison with reported experimental or theoretical data.
Note that we explicitly retain the $1/\omega^2$ factor, that compensates for using the time derivative $\dot{\alpha}(t)$ in the correlation function. 
The resulting lineshape function is a size-extensive spectral density of polarizability fluctuations with units of $\left[\alpha\right]^2 \times $ time.
It is further convenient to rewrite the equation in Cartesian components~\cite{McQuarrie2000, Xu-2024-JCTC}
\begin{equation}
    I_{\zeta\xi\sigma\tau}(\omega) 
    =
    \frac{1}{\omega^2}
    \frac{1}{2\pi} 
    \int_{-\infty}^\infty \, dt \, e^{-i\omega t}
    \left<
    \dot{\vec{\alpha}}_{\zeta\xi}(0) 
    \cdot 
    \dot{\vec{\alpha}}_{\sigma\tau}(t)
    \right>
    \, ,
    \label{eq:raman_lineshape_cartesian}
\end{equation}
where 
$\zeta$, $\xi$, $\sigma$, and $\tau$
are Cartesian components.
The resulting 4-d matrix containing $3^4$ elements can rigorously be mapped back to any arbitrary scattering geometry
\begin{equation}
    I'_\text{i,s}(\omega) 
    \propto
    \sum_{\zeta\xi\sigma\tau}
    \vec{\epsilon}_{\text{i,}\zeta} 
    \vec{\epsilon}_{\text{i,}\xi}
    \vec{\epsilon}_{\text{s,}\sigma}
    \vec{\epsilon}_{\text{s,}\tau}
    \cdot
    I_{\zeta\xi\sigma\tau}(\omega)
    \label{eq:raman_lineshape_is}
\end{equation}
where $\epsilon_i$ and $\epsilon_s$ are again the unit vectors of the polarization of the incident and scattered light beams, see above, and $\zeta$, $\xi$, $\sigma$, $\tau$ are their Cartesian components.
At this stage, \eref{eq:raman_lineshape_cartesian} enables us to condense all necessary information regarding Raman spectroscopy into 81 Cartesian spectra which can readily be computed in the (usually Cartesian) reference frame of a MD simulation, provided that polarizability time derivatives are available.
The number of spectra is further reduced by taking the symmetry $\alpha_{\zeta\xi} = \alpha_{\xi\zeta}$ into account which holds for most applications~\cite{Feynman-Lectures2}.
Consequently, we have 
$I_{\zeta\xi\sigma\tau}(\omega) = I_{\xi\zeta\sigma\tau}(\omega) = I_{\zeta\xi\tau\sigma}(\omega) = I_{\xi\zeta\tau\sigma}(\omega)$
reducing the number of independent Cartesian spectra to 21.
Using \eref{eq:raman_lineshape_is} any arbitrary scattering geometry can be synthesized from those 21 Cartesian spectra \emph{a posteriori}.
Therefore it is fully sufficient to store only the 21 independent components of 
$I_{\zeta\xi\sigma\tau}(\omega)$ 
to keep the full information on all recordable Raman spectra.
Even if different scattering geometries are to be calculated at a later stage, the MD simulations do not need to be revisited again, making 
$I_{\zeta\xi\sigma\tau}(\omega)$ 
an ideal output for long-term storage.

Having worked out the equations for any arbitrary system, it is important to note that the situation greatly simplifies for isotropic systems and isotropic experimental setups.
Following the derivations by Berne and Pecora~\cite{Berne-2000-DynamicLightScattering}, the polarizability tensor is decomposed into its scalar part $a(t)$ and its traceless anisotropic part $\beta(t)$, such that  $\alpha(t) = a(t) \vec{I} + \beta(t)$, where $a(t) = 1/3\,\, \text{Tr}\,[\alpha(t)]$, $\text{Tr}\,[\beta(t)] = 0$, and $\vec{I}$ is the unit tensor.
As alluded to previously, we use the Fourier transform identity to use the time derivative, $\dot{\vec{\alpha}}(t)$.
The Fourier-Transform identity can be applied to the published time correlation functions that use $\alpha(t)$ throughout~\cite{Berne-2000-DynamicLightScattering}, without requiring further alterations.
The isotropic spectrum is then obtained according to 
\begin{equation}
    I_\text{iso}(\omega) = 
    \frac{1}{\omega^2}
    \frac{1}{2\pi} 
    \int_{-\infty}^\infty \, dt \, e^{-i\omega t}
    \left<
    \dot{a}(0) 
    \cdot 
    \dot{a}(t)
    \right>
    =
    \frac{1}{9}
    \sum_{ij}
    I_{iijj}(\omega)
    \, .
    \label{eq:raman_lineshape_berne_iso}
\end{equation}
Further, the perpendicular (``VH'', $\perp$) correlation function is defined
\begin{equation}
    I_\text{VH}(\omega) = 
    \frac{1}{\omega^2}
    \frac{1}{2\pi} 
    \int_{-\infty}^\infty \, dt \, e^{-i\omega t}
    \frac{1}{10}
    \left<
    \text{Tr}
    \left[
    \dot{\beta}(0) 
    \cdot 
    \dot{\beta}(t)
    \right]
    \right>
    =
    \frac{1}{10}
    \sum_{ij}
    \left(
    I_{ijij}(\omega)
    -
    \frac{1}{3}
    I_{iijj}(\omega)
    \right)
    \, ,
    \label{eq:raman_lineshape_berne_vh}
\end{equation}
where the $1/10$ stems from the number of independent components in the traceless symmetric matrix $\beta(t)$.
$I_\text{VH}(\omega)$
can be combined with $I_\text{iso}(\omega)$ to yield the 
parallel (``VV'', $\parallel$)
spectrum,
\begin{equation}
    I_\text{VV}(\omega) = 
    I_\text{iso}(\omega) + 
    \frac{4}{3} I_\text{VH}(\omega)
    \, .
    \label{eq:raman_lineshape_berne_vv}
\end{equation}
Notably, also a forth , the ``anisotropic'', spectrum is commonly reported~\cite{Long-2002-TheRamanEffect}, 
\begin{equation}
    I_\text{VV}(\omega) = I_\text{iso}(\omega) + \frac{4}{45} I_\text{aniso}(\omega)
\end{equation}
that differs from $I_\text{VH}$ only by a constant prefactor of 15.

These spectra represent different physical aspects of the Raman scattering process.
The names ``VV'' and ``VH'' originate from the Porto notation, defining a parallel and perpendicular experimental setup, respectively. 
For the VV spectrum, scattered light with the same polarization as the incident light is detected, while for the perpendicular spectrum the polarization of the scattered light is perpendicular to the one of the incident light. 
In an isotropic system, the invariant properties are $a(t)$ and $\beta(t)$ as elaborated above, and the VV and VH spectra need to be converted to the isotropic and anisotropic 
kinds
that directly reflect the intrinsic isotropic and anisotropic components of the molecular polarizability fluctuations.
All four spectra are regularly reported in the literature.
Using the two equations \eref{eq:raman_lineshape_berne_iso} and \ref{eq:raman_lineshape_berne_vh}, we can now express all four relevant Raman spectra for isotropic systems, iso, aniso, VV, and VH, in terms of time correlation functions of the time derivative of the polarizability tensor in Cartesian components as defined in \eref{eq:raman_lineshape_cartesian}.

Notably, the VV and VH spectra are sometimes also referred to as ``polarized'' and ``depolarized'', respectively, referring to the polarization of the detected light. 
This terminology becomes clear considering the depolarization ratio~\cite{Long-2002-TheRamanEffect}, which is defined as $I_\text{VH}(\omega)/I_\text{VV}(\omega)$, i.e., the ratio of the ``depolarized'' to the ``polarized'' spectrum, and thus quantifies the extent to which the polarization of the incident light is lost upon scattering. 
This usage should not be confused with the molecular perspective, in which the isotropic and anisotropic spectra have also been termed ``polarized'' and ``depolarized'', respectively~\cite{McQuarrie2000}, where the distinction instead refers to the symmetry of the polarizability tensor. 
Care must therefore be taken when comparing with existing literature.

In the same spirit as for the IR spectrum, the chain rule can now be applied to the time derivative of the polarizability tensor
\begin{equation}
    \dot{\alpha}_{\zeta\xi} (t)
    = 
    \sum_{i,\eta} 
    \left.\frac{\partial \alpha_{\zeta\xi}}{\partial r_{i,\eta}}\right|_t 
    \left.\frac{\partial r_{i,\eta}}{\partial t}\right|_t
    \equiv 
    \sum_{i,\eta}
    \mathcal{Q}_{i, \eta, \zeta\xi}(t) \cdot v_{i, \eta}(t)
    \, ,
    \label{eq:adot_from_pgt}
\end{equation}
where we have introduced the atomic polarizability gradient tensor~(PGT), $\mathcal{Q}_{i,\eta,\zeta\xi}(t)$.
Note that the PGT is a rank-3 tensor rigorously defined for each atom $i$, where the index $\eta$ represents a Cartesian component of the atomic displacement.
We will see in \sref{sec:electronic_structure} that the indices $\zeta$ and $\xi$ represent Cartesian components of applied electric fields.
$\eta$ and $\zeta$ are therefore both consistent to our definition of the APT in \eref{eq:Mdot_from_apt}.
Inserting \eref{eq:adot_from_pgt} into \eref{eq:raman_lineshape_cartesian} enables one to represent all Cartesian Raman lineshapes and subsequently obtain all corresponding Raman spectra for any scattering geometry $I_{\zeta\xi\sigma\tau}(\omega)$ for non-isotropic and isotropic systems.
We stress that, in full analogy to the APT, $\mathcal{P}_i$, and the total dipole moment time derivative, this provides a rigorous atomic decomposition of the total polarizability time derivative, not relying on any artificial charge partitioning schemes.

\subsection{Quantifying Disagreements}

Having introduced APTs and PGTs as primary training targets in this work, it is necessary to quantify disagreement between APTs and PGTs when they are obtained from different sources. 
This is not only relevant for comparing \emph{ab initio} values with the ML predicted ones, but particularly also to compare different electronic structure approaches, see below.
Being the most common metric, we employ a component-wise RMSE, yielding the error for each component of $\mathcal{P}$ and $\mathcal{Q}$ individually. 
Since the RMSE is in the units of the property being either $\mathcal{P}$ or $\mathcal{Q}$, we further use a component-wise relative RMSE
\begin{equation}
    \delta^{\mathcal{Y}}
    =
    \frac{1}{N_\mathcal{Y}} \sum_{\eta, \chi}
    \delta^{\mathcal{Y}}_{\eta, \zeta}
    =
    \frac{1}{N_\mathcal{Y}} \sum_{\eta, \chi}
    \frac{
        \text{RMSE}_{\eta,\zeta}\left(
        \mathcal{Y}^\text{A}_{i, \eta, \chi},
        \mathcal{Y}^{\text{B}}_{i, \eta, \chi}
        \right)
    }
    {
        \sigma_{\eta,\zeta}\left(
        \mathcal{Y}^\text{B}_{i, \eta, \chi}
        \right)
    }
    \label{eq:apt_rel_rmse}
\end{equation}
which can be defined for APTs ($\mathcal{Y}=\mathcal{P}$, $N_\mathcal{P} = 9$, $\chi=\zeta$) and 
PGTs ($\mathcal{Y}=\mathcal{Q}$, $N_\mathcal{Q}= 27$, $\chi=\zeta\xi$).
For each component of $\mathcal{Y}$, the RMSE is computed between the corresponding components of set A and set B, where the two sets 
contain the very same configurations, but APTs and/or PGTs have been calculated in different ways.
Each component-wise RMSE is then normalized by the corresponding standard deviation $\sigma$, representing its 
natural variability
taken from data set B.
We then average the component-wise relative RMSEs to obtain a single value representing the mean relative deviation of set A with respect to the reference set B, yielding a dimensionless single number that characterizes the overall relative disagreement.
Previously, similar relative uncertainties were reported and transformed to a percentage score by dividing the force RMSE by the root mean square~\cite{Schran-2021-PNAS}. 
Herein, the standard deviation is used instead of the root mean square, because the mean of the individual components of $\mathcal{P}$ and $\mathcal{Q}$ is not necessarily zero.
If 
$\delta^\mathcal{Y} << 1$,
the absolute difference between
two  sets A and B is negligible compared to the spread expected from intrinsic fluctuations.
The relative RMSE can further directly be transferred into a mean \emph{Coefficient of Determination,} 
\begin{equation}
    R^2_\mathcal{Y} = 
    \frac{1}{N_\mathcal{Y}} \sum_{\eta, \chi}
    \left[
    1 - 
    \left(
    \delta^\mathcal{Y}_{\eta,\chi}
    \right)^2
    \right]
    \, ,
    \label{eq:R2}
\end{equation}
which is a common metric for fitting data to models
and can be interpreted as the ``explained variance''.
For instance, a value of $R^2=0.95$ indicates that the model explains 95~\% of the variance in the underlying test set.
Importantly, we compute $R^2$ for each component separately and only then average over all $R^2$ values to yield a mean overall \emph{Coefficient of Determination}.

It should be noted that none of these metrics are universal and it strongly depends on the context what ``good'' RMSEs, relative RMSEs, or $R^2$ coefficients are. 
In the following we will address these metrics in terms of the conversion of the underlying spectra, i.e.\ what metrics are necessary to result in converged spectra. 
Recall that the target property that we are actually interested in are vibrational spectra, thus their convergence with respect to the corresponding true \emph{ab initio} spectra is the relevant quality measure. 
We quantify the difference between two spectra as
\begin{equation}
    \Delta^{I(\omega)} = 
    \frac{
        \int_0^\infty \left| I^\text{A}(\omega) - I^\text{B}(\omega) \right| d\omega 
    }
    {
        \int_0^\infty \left| I^\text{A}(\omega) \right| d\omega 
        +
        \int_0^\infty \left| I^\text{B}(\omega) \right| d\omega 
    }
    \label{eq:lineshape_difference}
\end{equation}
where 
$I^\text{A/B} (\omega)$ is the lineshape function of a IR (\eref{eq:ir_from_Mdot}) or Raman~(\eref{eq:raman_lineshape_cartesian}) spectrum, 
where the A and B spectra were obtained for the same trajectories, but the APTs and/or PGTs were sourced differently.
Note that this score is similar to the one introduced previously for 
radial distribution functions and
vibrational density of states~\cite{Schran-2021-PNAS}.
Here we use absolute values for the spectra in the denominator as well, since cross correlation spectra, see below, can also be negative.

\subsection{Calculating APTs and PGTs from Electronic Structure Theory}
\label{sec:electronic_structure}

By definition, APTs and PGTs are most naturally obtained by directly evaluating the spatial derivatives of the dipole moment and the polarizability tensor with respect to atomic displacements.
This can be done numerically by manually displacing each atom individually for a given configuration.
For APTs, an analytical route based on density functional perturbation theory is also available~\cite{Ditler-2021-JCP}.
The latter approach, however, still requires the evaluation of the APT for each atom individually, such that the computational effort scales with the number of atoms $N$ in both cases.

A different approach 
can be obtained when expanding the total energy of the system as a function of an applied electric field~\cite{SzaboOstlund}
\begin{equation}
    E(\vec{\field}) = E(0) + 
    \sum_\zeta \left( \frac{\partial E(\vec{\field})}{\partial \field_\zeta} \right)_0 
    \field_\zeta
    + 
    \frac{1}{2}
    \sum_{\zeta\xi} 
    \left( \frac{\partial^2 E(\vec{\field})}{\partial \field_\zeta \field_\xi} \right)_0
    \field_\zeta\field_\xi
    + 
    \cdots
    \, ,
\end{equation}
where $\field_\zeta$ is the $\zeta$-th Cartesian component of the field vector and we identify~\cite{SzaboOstlund}
\begin{equation}
    E(\vec{\field}) = E(0) - 
    \sum_\zeta 
    M_\zeta
    \field_\zeta
    - 
    \frac{1}{2}
    \sum_{\zeta\xi} 
    \alpha_{\zeta\xi}
    \field_\zeta\field_\xi
    + 
    \cdots
    \, ,
\end{equation}
where $M_\zeta$ is the $\zeta$-th component of the dipole moment and $\alpha_{\zeta\xi}$ is the $\zeta\xi$-th component of the polarizability tensor.
Note that the next following term is the hyperpolarizability.
Taking the 
derivative
along the $\eta$-th cartesian component 
of atom $i$ 
we can rewrite the energy expansion above
\begin{equation}
    F_{i,\eta}(\vec{\field}) = 
    - \frac{\partial E(0)}{\partial r_{i,\eta}} 
    + 
    \sum_{\zeta} \frac{\partial M_\zeta}{\partial r_{i, \eta}} 
    \field_\zeta
    +
    \frac{1}{2}
    \sum_{\zeta\xi}
    \frac{\partial \alpha_{\zeta\xi}}{\partial r_{i, \eta}} 
    \field_\zeta \field_\xi
    + 
    \cdots
    \, , 
\end{equation}
where $F_{i,\eta}(\field)$ is the force acting on atom $i$ along $\eta$ in an external electric field $\vec{\field}$.
Expanding that force as a function of an applied electric field in analogy to the energy expansion above,
\begin{equation}
    F_{i,\eta}(\vec{\field}) = 
    F_{i,\eta}(0)
    +
    \sum_\zeta \left( \frac{\partial F_{i, \eta}(\vec{\field})}{\partial \field_\zeta} \right)_0 
    \field_\zeta
    + 
    \frac{1}{2}
    \sum_{\zeta\xi} 
    \left( \frac{\partial^2 F_{i,\eta}(\vec{\field})}{\partial \field_\zeta \field_\xi} \right)_0
    \field_\zeta\field_\xi
    + 
    \cdots
    \label{eq:force-expansion}
\end{equation}
we 
extract the identities
\begin{equation}
    \mathcal{P}_{i, \eta, \zeta} =
    \frac{\partial M_\zeta}{\partial r_{i, \eta}} =
    \frac{\partial F_{i, \eta}}{\partial \field_\zeta}
    \label{eq:apt-identity}
\end{equation}
for the APT
and
\begin{equation}
    \mathcal{Q}_{i, \eta, \zeta\xi} = 
    \frac{\partial \alpha_{\zeta\xi}}{\partial r_{i, \eta}} = 
    \frac{\partial^2 F_{i, \eta}}{\partial \field_\zeta \partial \field_\xi}
    \label{eq:pgt-identity}
\end{equation}
for the PGT.
Note that we assume that the field is homogeneous~\cite{Jackson1975, SzaboOstlund} and that \eref{eq:force-expansion} has been used recently to represent external electric fields in MLMD simulations in a perturbative way~\cite{Joll-2024-NatCommun}.
These identities are critical because forces are very efficiently available from DFT codes for all atoms in a given configuration simultaneously, thereby lifting the unfavorable scaling with the number of atoms associated with explicit spatial displacements.
Note that spatial and field derivatives have regularly been used in the past for APTs in the context of ML, see, e.g., Refs.~\cite{Schienbein-2023-JCTC, Schmiedmayer-2024-JCP, Joll-2024-NatCommun, Jindal-2025-JCTC}, while PGTs have not been employed so far.
Moreover, these identities provide a powerful consistency check for the numerical accuracy of the calculated APTs and PGTs, since the different derivatives should, in principle, yield identical results.

Numerically calculating the first derivatives, $\frac{\partial M_\zeta}{\partial r_{i, \eta}}$, $\frac{\partial F_{i, \eta}}{\partial \field_\zeta}$, and $\frac{\partial \alpha_{\zeta\xi}}{\partial r_{i, \eta}}$ is straightforward through a central finite difference.
Some more attention requires the second numerical derivative of the of the force with respect to two electric fields in \eref{eq:pgt-identity}.
First, we define two finite differences
\begin{equation}
    \Delta S_{i, \eta, \zeta} = 
    \frac{ 
    F_{i, \eta}(\field_\zeta = +h) - 2 F_{i,\eta}(0) + F_{i,\eta} (\field_\zeta = -h)
    }
    {h^2}
    \, ,
    \label{eq:pgt-numderiv-S}
\end{equation}
and
\begin{equation}
    \Delta D_{i, \eta, \zeta\xi} = 
    \frac{
        F_{i, \eta}(\field_\zeta = \field_\xi = h) 
        - 
        2 F_{i, \eta}(0)
        +
        F_{i, \eta}(\field_\zeta = \field_\xi = -h) 
    }
    {h^2}
    \label{eq:pgt-numderiv-D}
\end{equation}
where $h$ is the step size of the numerical difference as usual.
Note that the second equation is a standard numerical second derivative, but taken along a mixed line along $\zeta$ and $\xi$.
Using those two finite differences, we can express the diagonals of $\mathcal{Q}$
\begin{equation}
    \mathcal{Q}_{i, \eta, \zeta\zeta} = 
    \Delta S_{i,\eta,\zeta}
    \label{eq:pgt-numderiv-diag}
\end{equation}
and the off-diagonals
\begin{equation}
    \mathcal{Q}_{i, \eta, \zeta\xi} = \frac{1}{2} 
    \left( \Delta D_{i, \eta, \zeta\xi} - 
    \Delta S_{i, \eta, \zeta} - \Delta S_{i, \eta, \xi} 
    \right)
    \, , \, \zeta\neq\xi
    \, ,
    \label{eq:pgt-numderiv-offdiag}
\end{equation}
where 
the symmetry of the field derivatives,
$\frac{\partial^2 F_{i,\eta}}{\partial \field_\zeta \partial \field_\xi} = \frac{\partial^2 F_{i,\eta}}{\partial \field_\xi \partial\field_\zeta}$, 
is
exploited.
Using these two equations requires 13 single point calculations in total to calculate \emph{all} PGTs of a given configuration, 
instead of 21 if the second derivative was not symmetric.
Importantly, six of these single point calculations, 
namely $F_{i,\eta}(\field_\zeta = \pm h)\,\forall\zeta$,
are also required when calculating the APT using the electric field derivative.
This implies that calculations used for obtaining APTs can be recycled for calculating PGTs and vice-versa.
Consequently, all APTs and PGTs for a given configuration can be obtained from a total of 13 single-point calculations.

\subsection{Automated training of APTs and PGTs}
\label{sec:training}
Following our previous work introducing APTs as ML targets~\cite{Schienbein-2023-JCTC}, we here extend the concept by introducing PGTs as new direct ML targets for predicting accurate Raman 
spectra.
For the training, we employ the  \texttt{e3nn} framework~\cite{e3nn} built on top of \texttt{pytorch}~\cite{pytorch}.
\begin{comment}
    Later, 
    various ML architectures 
    have been used 
    to predict APTs, 
    including MACE~\cite{Kapil-2024-Faraday, Stocco-2025-npj}, Nequip~\cite{Bergmann-2025-PRL}, or Gaussian Process Regression~\cite{Schmiedmayer-2024-JCP}.
    In contrast to these approaches, our philosophy is to train APTs and PGTs \emph{directly}, rather than training the corresponding global properties~--
    namely the dipole moment (for APTs) and the polarizability (for PGTs)~-- and then obtaining their derivatives, for instance 
    via automatic differentiation.
    Both training strategies have been demonstrated to achieve comparable accuracy, but they emphasize on different aspects of the underlying physics.
    Here, we exploit the fact that the APT and PGT tensors themselves are local and gauge- and branch-invariant response quantities. 
    This makes it possible to use gas-phase cluster data to reproduce IR spectra of periodic systems, as recently demonstrated~\cite{Jindal-2025-JCTC}.
    Conceptually, the multi-valuedness of the dipole moment in periodic systems~\cite{Spaldin-2012-JSolidStateChem} and the need to explicitly account for the total charge in systems with a non-zero net charge~\cite{Stocco-2025-npj} are avoided by construction in the direct tensor-training approach.
\end{comment}
Within the graph neural network scheme in \texttt{e3nn} each atom is represented by a node, and edges are constructed between atoms within a radial cutoff of \SI{6}{\angstrom}.
The input node features consist of a one-hot scalar encoding the chemical species ($0e$ irrep). 
For each edge, we compute a spherical-harmonic expansion up to $l_\text{max} =3$ and combine it with 10 radial basis functions of type \emph{smooth\_finite}.
Both the APTNN and the 
polarizability gradient tensor neural network~(PGTNN)
models employ three message-passing layers based on the \texttt{MessagePassing} class from \texttt{e3nn.nn.models.v2106.gate\_points\_message\_passing}.
Each message passing layer includes the full set of irreducible representations from $l=0$ to $l=3$ with even and odd parity. 
The multiplicities are (20, 10, 6, 5) channels for $l=0, 1, 2, 3$ in the APT model and (40, 20, 13, 10) channels, respectively, in the PGT model.
The number of channels therefore decreases with increasing $l$.
The output irreducible representations are determined directly by the tensorial structure of the target property.
The target tensors are transformed into their irreducible representations and standardized component-wise.
During prediction the inverse transformation is applied to recover the physical tensor components. 
For the optimization we use the \texttt{Adam} optimizer~\cite{Adam-optimizer}.
The PGTNN presented here constitutes the first implementation of direct training of a polarizability gradient tensor.
In addition, the APTNN has been completely reworked relative to our previous implementation~\cite{Schienbein-2023-JCTC} to be more memory- and runtime-efficient and substantially easier to use.
These architectural settings provide an efficient default configuration in terms of memory and GPU runtime, but can be overridden by the user if necessary.

While the code exposes all typical low-end routines for training and prediction, we also have an automated training script available:
We begin with a trained ML potential 
and run a 200~ps NVT simulation, from which we sample 80 independent configurations to subsequently spawn 80 times 20~ps NVE simulations.
From those 80 NVE trajectories, the IR and Raman spectra are to be calculated.
This NVT/NVE split has been established previously and used numerous times to get accurate vibrational spectra from \emph{ab initio} and MLMD simulations~\cite{Schienbein-2017-JPCL, Schienbein-2020-ANIE, Schienbein-2023-JCTC}.
The number of NVE trajectories can be adjusted depending on the statistical quality needed.
At this stage, the 80 NVE trajectories deliver positions and velocities as usual, but are still lacking 
the electronic information required for IR and Raman spectra.

To train an APTNN and/or PGTNN that can predict the required response tensors in a post-processing step, we have implemented an automated training procedure that hides most of the technical complexity from the user. 
The autotrainer randomly samples configurations from the 80 NVE trajectories to construct training and test sets. 
It also includes a built-in calculator that uses \texttt{CP2k} to numerically evaluate APTs or PGTs, for which the user must only supply the electronic-structure input. 
Alternatively, APTs and/or PGTs may be provided manually~-- for instance, if 
a different electronic structure code was used to train the ML potential.
The autotrainer then iteratively trains the APTs and PGTs until convergence, thereby minimizing the number of explicit electronic-structure calculations. 
For all test systems presented here, we sampled 200 and 100 configurations for the training and test sets, respectively. 
We emphasize that 200 training configurations are generous: in practice, convergence is usually achieved with substantially fewer samples, see below. 
Although we use random sampling for the training set generation, we employ an iterative training procedure, where the training set size is systematically increased.
We will see in the following that RMSEs and spectra systematically converge as a function of the training set size.

In the following, we consider two interatomic ML potentials previously reported in the literature:
bulk liquid water~\cite{Schienbein-2023-JCTC, Joll-2024-NatCommun} (RPBE-D3), 
and \ce{SO4^{2-}} (BLYP-D3) solutions~\cite{Schran-2021-PNAS}.
The aqueous \ce{SO4^{2-}} solution introduces the molecular ion as a rarely occurring species whose spectral response can be dominated by the water molecules in the total spectra.
The underlying potentials were trained using the RPBE~\cite{Hammer-1999-PRB} and BLYP~\cite{Gill-1992-ChemPhysLett} functionals for pure liquid water and the aqueous \ce{SO4^{2-}} solution, both 
supplemented by dispersion corrections (D3~\cite{Grimme-2010-JCP}).
We use the \texttt{CP2k} program code~\cite{VandeVondele-2005-ComputPhysCommun, Kuehne-2020-JCP} for all electronic structure calculations that are necessary to generate training and test data.
To be consistent to the trained ML potential, we employ precisely the same electronic structure setup that has been used for training the ML potential and we refer to the respective paper for the computational details.
For each system, we employ the training procedure outlined above and compute IR and Raman spectra.

As a benchmark, we explicitly compute \emph{ab initio} response tensors for \emph{one} full NVE trajectory per system.
The resulting spectra, obtained by calculating explicit \emph{ab initio} 
APTs and PGTs
along a ML potential-generated trajectory, are of course not statistically converged.
However, this single trajectory  allows us to assess the accuracy of the trained 
APTNN and PGTNN at the level of the relevant observable, namely the vibrational spectrum of interest. 
By using the same dynamics (positions and velocities), but APTs and PGTs sourced from either explicit \emph{ab initio} calculations or from the trained ML model
we fully eliminate any
contribution from errors in the underlying ML potential, in contrast to comparisons with fully explicit \emph{ab initio} MD.

\subsection{Cross Correlation Analysis}
\label{sec:cca}

Our goal is to benchmark the trained response models (APTNN, PGTNN) directly on the observable of interest, i.e., vibrational spectra.
Standard atom-wise metrics such as the RMSE are useful diagnostics, but they do not quantify 
directly 
translate into spectral inaccuracies. 
This becomes critical when rare structural motifs, such as 
\ce{SO4^{2-}},
contribute only weakly to the \emph{total} spectrum. 
For the low ion concentrations considered here, cleanly resolving solute 
signatures in the total spectrum in the sense of \emph{difference spectra}~\cite{Schmidt-JACS-2009, Schienbein-2017-JPCL}
would require many nanoseconds of trajectory data to achieve statistical convergence~\cite{Kann-2016-JCP}.
While obtaining such long trajectories and thus difference spectra is feasible with ML-based simulations, our benchmark must reference explicit \emph{ab initio} MD simulations.
At the \emph{ab initio} MD level, generating converged difference spectra is, however, not practical at all~\cite{Schienbein-2017-JPCL}.
A further complication is that water molecules in the vicinity of those rare structural motifs usually show a different spectral response than bulk water molecules.
Those water molecules can furthermore exchange in time since they continuously move between first and second solvation shells or between interfacial and bulk-like environments. 
Consequently comparing spectra ``per atom fixed in time'' only reproduces an average picture and does not specifically probe those atoms when they are in the vicinity of the ion or the interface.

To address these issues the \emph{Cross Correlation Analysis}~(CCA) was introduced some time ago for IR spectra~\cite{Schienbein-2017-JPCL}, which provides a generalized and systematic framework to decompose a total spectrum into topological- or motif-resolved contributions (e.g.\ first shell waters around ions or hydrogen bonds between first and second shells).
Crucially, CCA uses time-dependent indicator functions so that the membership in such a topological motif is dynamic.
The method guarantees exact additivity such that the total IR spectrum is recovered as the sum of all motif spectra, with no artificial loss or addition of intensity.
Using APTs, this framework extends to an ``atomic CCA''~\cite{Joll-2025-JPCL-THz}
\begin{equation}
    \left<\vec{\dot{M}}(0) \vec{\dot{M}}(t)\right> 
    \equiv
    \sum_{ab}
    C_{ab}^\gamma(t) = 
    \sum_{ab}
    \left<
    \sum_{i,j} \left(
    G_{a,i}(0)G_{b,j}(t) R_{\gamma,i,j}(0)
    \dot{\vec{\mu}}_i(0) 
    \dot{\vec{\mu}}_j(t)
    \right)
    \right>
    \, ,
    \label{eq:cca_ir_cf}
\end{equation}
where 
the total dipole moment time correlation function is decomposed in terms of groups and relations; 
$G_{a, i}(t)$ is the time-dependent group selector being unity if molecule $i$ is in group $a$ at time $t$
and
$R_{\gamma, i, j}(t)$ is the time-dependent relation selector being unity if molecules $i$ and $j$ are fulfill the relation $\gamma$ at time $t$.
We refer the reader to Ref.~\cite{Schienbein-2017-JPCL} for an elaborate discussion of groups and relations.
This equation was originally formulated in terms of molecules, where $\dot{\mu}$ was obtained from assigning Wannier centers to molecules.
The indices $i$ and $j$ can readily be reinterpreted as atom indices, 
as 
$\dot{\vec{\mu}}_i(t) = \mathcal{P}_i(t) \cdot v_i(t)$ 
(see \eref{eq:Mdot_from_apt})
and the atomic CCA~\cite{Joll-2025-JPCL-THz} is recovered. 
Inserting that correlation function into our definition of the IR absorption coefficient, \eref{eq:ir_from_Mdot}, we can readily express the total absorption coefficient of IR spectroscopy as the sum of all individual contributions from each group and relation
\begin{equation}
    \alpha(\omega) = 
    \sum_{ab\gamma}
    \alpha_{ab}^{\gamma}(\omega) = 
    \sum_{ab\gamma}
    \frac{\pi\beta}{3 V c \epsilon_0 n(\omega)} \frac{1}{2\pi} 
    \int_{-\infty}^\infty \, dt \, e^{-i\omega t}
    C_{ab}^{\gamma}(t)
    \, .
    \label{eq:cca_ir_lineshape}
\end{equation}

We now transfer the CCA concept to dissect Raman spectra as well.
Previously we have worked out that any arbitrary scattering geometry of Raman spectra can be expressed in terms of Cartesian lineshape functions, see \eref{eq:raman_lineshape_cartesian}.
We assume that the scattering geometries remain fixed in time for the duration of the NVE trajectory length, being 20~ps here.
We therefore introduce the short-hand notation 
$\widetilde{x} = \zeta\xi\sigma\tau$
for the scattering geometry, see \eref{eq:raman_lineshape_cartesian}, and perform one CCA analysis for each of the 21~unique Raman
lineshape functions.
Recall that this number 
reduces further for isotropic systems, see above.
To derive the atomic Raman CCA, we insert \eref{eq:adot_from_pgt} into \eref{eq:raman_lineshape_cartesian} to rewrite the time correlation function as follows
\begin{equation}
    \left<
    \dot{\vec{\alpha}}_{\zeta\xi}(0) 
    \dot{\vec{\alpha}}_{\sigma\tau}(t)
    \right>
    =
    \left<
    \sum_{ij}
    \dot{\alpha}^{(i)}_{\zeta\xi}(0)
    \dot{\alpha}^{(j)}_{\sigma\tau}(t)
    \right>
    \, ,
\end{equation}
where we introduced the short-hand notation 
\begin{equation}
    \dot{\alpha}^{(i)}_{\zeta\xi} (t) = \mathcal{Q}_{i,\eta,\zeta\xi}(t) v_{i,\eta}(t)
\end{equation}
to represent the change of the total polarizability $\dot{\alpha}(t)$ along the two Cartesian components $\zeta$ and $\xi$ at time $t$ due to atom $i$.
Note that index $\eta$ is summed over.
Inserting the definitions for groups and relations results in
\begin{equation}
    \left<
    \dot{\vec{\alpha}}_{\zeta\xi}(0) 
    \dot{\vec{\alpha}}_{\sigma\tau}(t)
    \right>
    =
    \sum_{ab}
    \left<
    \sum_{ij}
    G_{a,i}(0)G_{b,j}(t) R_{\gamma,i,j}(0)
    \dot{\alpha}^{(i)}_{\zeta\xi}(0)
    \dot{\alpha}^{(j)}_{\sigma\tau}(t)
    \right>
    \equiv
    \sum_{ab}
    \left(C_{\widetilde{x}}\right)_{ab}^\gamma(\omega)
    \, ,
    \label{eq:cca_raman_lineshape}
\end{equation}
where 
the time-dependent group selectors $G_{a,i}(t)$ and relation selectors $R_{\gamma, i, j}(t)$ are exactly the same as for the IR case.
The obtained correlation function can now be inserted into the equation for the lineshape function, \eref{eq:raman_lineshape_cartesian}, yielding
\begin{equation}
    I_{\widetilde{x}}(\omega) 
    =
    \sum_{ab\gamma}
    \left(I_{\widetilde{x}}\right)_{ab}^\gamma(\omega)
    = 
    \sum_{ab\gamma}
    \frac{1}{\omega^2}\frac{1}{2\pi} 
    \int_{-\infty}^{\infty}
    dt\,
    e^{-i\omega t}
    \left(C_{\widetilde{x}}\right)_{ab}^\gamma(\omega)
    \, .
    \label{cca_raman_lineshape}
\end{equation}
Note that a CCA decomposition can result in a large number of individual spectra which is inflated for Raman spectroscopy due to the different scattering geometries. 
For benchmarking purposes it is, however, not necessary to manually check each and every spectrum:
We apply the atomic Raman CCA consistently to the reference \emph{ab initio} trajectory and the ML trajectory and then compare \emph{all} individual spectra using \eref{eq:lineshape_difference}.
Thereby, we can systematically assess how well each topological motif performs and, most importantly, if rare species like
\ce{SO4^{2-}} ions,
are equally well represented as the bulk water molecules which 
are much more abundant.
For the aqueous solutions inspected herein, we employ a CCA decomposition based on the solute, the first and second shell water molecules, and all water molecules beyond the second shell.
That decomposition turned out to be successful for aqueous ion solutions, showing markedly different spectral contributions of water molecules in these three groups~\cite{Schienbein-2017-JPCL}.

\section{Results and Discussion}

\subsection{Numerical Derivatives from Electronic Structure Theory}

\begin{figure}
    \begin{subfigure}{0.49\textwidth}
        \includegraphics[width=\textwidth]{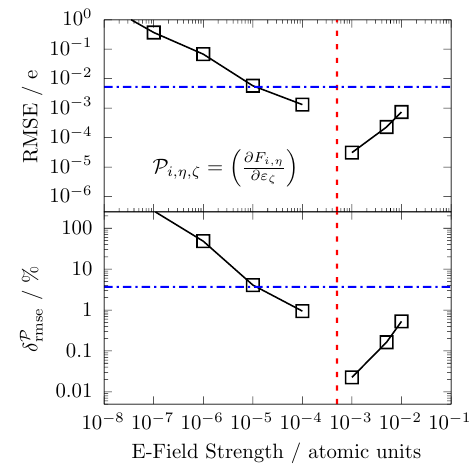}
        \caption{}
        \label{fig:apt-efield-numderiv}
    \end{subfigure}
    \hfill
    \begin{subfigure}{0.49\textwidth}
        \includegraphics[width=\textwidth]{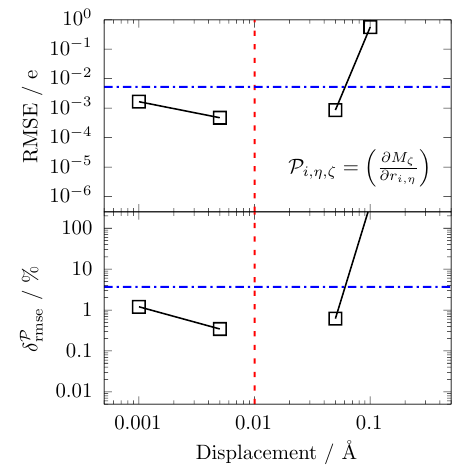}
        \caption{}
        \label{fig:apt-spatial-numderiv}
    \end{subfigure}
    \caption{
        RMSE (top panels) and relative RMSE, $\delta_\text{rmse}^\mathcal{P}$, see \eref{eq:apt_rel_rmse} (bottom panels), 
        when computing APTs from taking the derivative of atomic forces with respect to an electric field (a), \eref{eq:apt-identity}, last term,
        or from taking the derivative of the total dipole moment with respect to an atomic displacement in space (b), \eref{eq:apt-identity}, central term,
        both evaluated numerically by central finite differences using the field strength or displacement given at the respective abscissa. 
        The electric field test is based on 30 configurations, while the spatial derivative test is based on a single configuration, all containing 128 water molecules each, see text.
        The vertical red dashed line indicates the respective reference value, being \num{5e-4} atomic units ($\approx \SI{0.026}{\VpA}$) in case of the electric field derivative and 
        \SI{0.01}{\angstrom} in case of the spatial derivative.
        The horizontal blue dashed-dotted line indicates the respective deviations when comparing the field with the spatial derivative, see \fref{fig:apt-dft-rmse-comparison}.
    }
    \label{fig:apt-numderiv}
\end{figure}

\begin{figure}
    \begin{subfigure}{0.49\textwidth}
        \includegraphics[width=\textwidth]{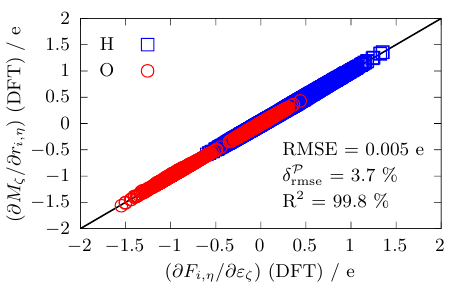}
        \caption{}
        \label{fig:apt-dft-rmse-comparison}
    \end{subfigure}
    \hfill
    \begin{subfigure}{0.49\textwidth}
        \includegraphics[width=\textwidth]{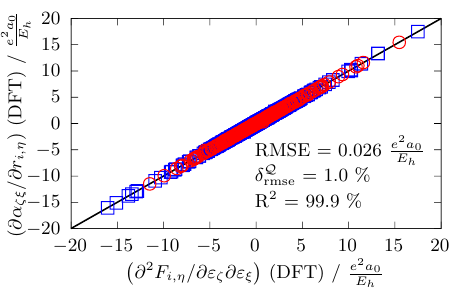}
        \caption{}
        \label{fig:pgt-dft-rmse-comparison}
    \end{subfigure}
    \caption{
        Parity plots comparing APTs (a) and PGTs (b) obtained from numerical electric field derivatives (abscissa) and spatial derivatives (ordinate), respectively. 
        According to the identities in \eref{eq:apt-identity} and \eref{eq:pgt-identity} they should be identical, the parity plot visualizes the deviation and the printed RMSE, relative RMSE, and R$^2$ values quantify the disagreement. 
        The test is performed for a single configuration of liquid water, containing 128 water molecules. 
    }
\end{figure}

Calculating the numerical derivatives at the electronic structure level
requires carefully choosing the displacement of each atom for spatial derivatives or the field strength for the electric field derivative.
For that reason, we screen field strengths and spatial displacements in the following.
We begin by taking 30 configurations containing 128 water molecules and compute APTs for all atoms in all configurations employing the electric field derivative,
requiring in total 180 electronic structure calculations (6 per configuration) for each screened electric field strength.
In contrast, to obtain all APTs for all 128 water molecules in a single configuration by a spatial derivative we require \num{2304} electronic structure calculations (6 per atom).
Due to that computational cost,
we perform the spatial displacement screening for a single configuration only.
Simultaneously the cost illustrates the efficiency boost when using field derivatives.
The field and displacement screenings are depicted in \fref{fig:apt-numderiv}a and b, respectively.
For the obtained APTs we then compute the RMSE and $\delta_\text{rmse}^\mathcal{P}$ recovering the expected U-shaped behavior of numerical derivatives.
We identify the minimum at around \num{5e-4}~au and \SI{0.01}{\angstrom}, respectively, confirming the choices made in previous papers~\cite{Schienbein-2023-JCTC, Joll-2024-NatCommun}.
Taking these two optimal values, we compare the obtained APTs between the spatial and the field derivative in \fref{fig:apt-dft-rmse-comparison}.
Visually, the agreement between the two is excellent, more interesting is the quantitative disagreement given by the RMSE (0.005~e), the relative RMSE ($\delta_\text{rmse}^\mathcal{P}=3.7\,\%$), and the R$^2$ value (99.8~\%).

\begin{figure}
    \begin{subfigure}{0.49\textwidth}
        \includegraphics[width=\textwidth]{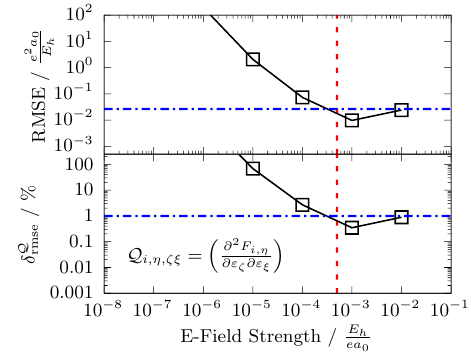}
        \caption{}
        \label{fig:pgt-efield-numderiv}
    \end{subfigure}
    \hfill
    \begin{subfigure}{0.49\textwidth}
        \includegraphics[width=\textwidth]{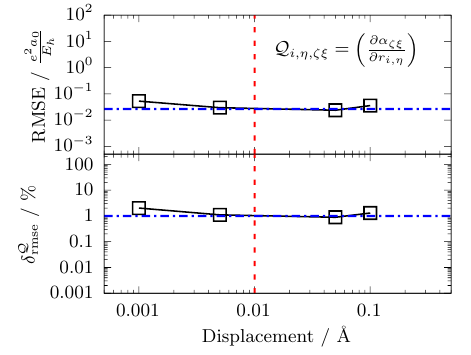}
        \caption{}
        \label{fig:pgt-spatial-numderiv}
    \end{subfigure}
    \caption{
        RMSE (top panels) and relative RMSE, $\delta_\text{rmse}^\mathcal{Q}$, see \eref{eq:apt_rel_rmse} (bottom panels), 
        when computing PGTs from taking the derivative of atomic forces with respect to an electric field (a), \eref{eq:pgt-identity}, last term,
        or from taking the derivative of the polarizability tensor with respect to an atomic displacement in space (b), \eref{eq:pgt-identity}, central term,
        both evaluated numerically by central finite differences using the displacement given at the respective abscissa.
        The electric field test is based on 30 configurations, while the spatial derivative test is based on a single configuration, all containing 128 water molecules each, see text.
        The vertical red dashed line indicates the respective reference value, being \num{5e-4} atomic units ($\approx \SI{0.026}{\VpA}$) in case of the electric field derivative and 
        \SI{0.01}{\angstrom} in case of the spatial derivative.
        The horizontal blue dashed-dotted line indicates the respective deviations when comparing the field with the spatial derivative, see \fref{fig:pgt-dft-rmse-comparison}.
    }
    \label{fig:pgt-numderiv}
\end{figure}

The same exercise is repeated for the PGT to determine suitable displacement values for the corresponding numerical derivatives as well. 
We calculate PGTs via finite field derivatives through \eref{eq:pgt-numderiv-diag} and \ref{eq:pgt-numderiv-offdiag} for 30 configurations of liquid water, containing 128 water molecules each, requiring in total 390 single point calculations.
Since the quadratic scaling of the applied electric field strength in the denominator of the second derivative
we tightened the SCF calculation significantly to a value of \num{1e-9} for \texttt{EPS\_SCF} which is generous, see discussion below.
For the spatial derivative we again displace each atom individually in space by a small distance in analogy to the APT, where we employ density functional perturbation theory in \texttt{CP2k} to obtain the polarizability tensor~\cite{Putrino-2002-PRL, Luber-2014-JCP, Kuehne-2020-JCP} at each displaced configuration.
Here we compute PGTs for a single configuration containing 128 water molecules, requiring in total 2304 DFPT calculations. 
The field and displacement screenings of the PGT are depicted in \fref{fig:pgt-numderiv}a and b, respectively.
We take the same field strength (\num{5e-4}~au) and spatial displacement (\SI{0.01}{\angstrom}) as reference, indicated by the red dashed vertical line.
We plot the RMSE and $\delta_\text{rmse}^\mathcal{Q}$, 
compared to these reference values
observing the typical U-shaped behavior.
We identify a quite steep increase of the error for smaller fields than \num{1e-4}~au since the field strength enters quadratically through the denominator of the numerical derivative, \eref{eq:pgt-numderiv-S} and \ref{eq:pgt-numderiv-D}.
In \fref{fig:pgt-dft-rmse-comparison} we present the parity plot comparing PGTs obtained from field and spatial derivatives.
Overall we find good visual agreement, resulting in a RMSE of $0.026~\frac{e^2a_0}{E_h}$, a relative RMSE of \SI{1.0}{\percent}, and a $R^2$ coefficient of \SI{99.9}{\percent}.

We have computed APTs and PGTs by two different approaches using the very same electronic structure setup. 
Since the two approaches are equivalent, the residual disagreement reflects an intrinsic
numerical uncertainty of APTs and PGTs inherent for the chosen electronic structure setup.
The herein employed DFT settings are representative of a well-converged and numerically stable \emph{ab initio} MD setup, as demonstrated for the same system in previous work~\cite{Imoto-2015-PCCP, Schienbein-2018-JPCB, Schienbein-2020-PCCP, Schienbein-2020-ANIE, Mauelshagen-2025-SciAdv}.
At the same time, the remaining discrepancy indicates that achieving even closer agreement between spatial and field derivatives would likely require tighter energy convergence~-- not tighter SCF precision~-- with respect to several parameters at the electronic structure level, for instance the basis set size.
Consequently, all field strengths and atomic displacements that yield errors below this intrinsic uncertainty, corresponding to all data points below the blue horizontal line in \fref{fig:apt-numderiv}a and b, and \fref{fig:pgt-numderiv}a and b, are equally suitable choices for computing numerical finite differences.
This provides a wide range of admissible displacements that lead to equally accurate APTs.
In case of PGTs, the range for accurate field derivatives is, however, significantly smaller.

By comparing PGTs computed from two independent sources, the numerically less stable second electric field derivative is validated against the spatial first derivative, showing excellent agreement ($R^2=\SI{99.9}{\percent}$) and thereby justifying the approach.
Still, we observe a quite strong dependence of the PGT with respect to the applied field strength (\fref{fig:pgt-efield-numderiv}) that is much stronger than for the APT (\fref{fig:apt-efield-numderiv}).
That behavior is understandable in view of the quadratic denominator in case of the PGT.
We present the convergence of the PGT with respect to the SCF precision in \fref{fig_si:pgt_eps} in the SI.
The figure shows that the relative error 
is smaller than \SI{4}{\percent} ($R^2=\SI{99.8}{\percent}$)
between \texttt{EPS\_SCF} values of \num{1e-8} and \num{1e-10}.
This implies that the herein consistently applied convergence criterion of \num{1e-9} is more than sufficient.
Generally, we therefore recommend a SCF convergence criterion smaller than \num{1e-8} for \texttt{CP2k} and emphasize that this should be tested for different electronic structure codes and might also be system or basis set dependent. 
However, the tight SCF convergence is primarily to achieve high precision in the calculated forces.
That is necessary to compensate for the division by $h^2$ in \eref{eq:pgt-numderiv-S} and \ref{eq:pgt-numderiv-D}.
Given the 
narrow window
for electric field values to obtain accurate PGTs, see above, and the required tightness of the SCF calculation, we generally recommend to do a benchmark of the numerical derivative for these two parameters prior training a PGTNN as illustrated herein.
Finally it should be noted that the tight convergence of the SCF and the additional benchmarks add some overhead for the calculation of PGTs.
Nevertheless, obtaining PGTs by electric field derivatives is still much more 
economical than using spatial derivatives because the former only requires 6 single point calculations for a full configuration, compared to $6N$ DFPT calculations in case of the latter.

\subsection{Convergence of the APTNN}

\begin{figure}
    \centering
        \includegraphics[width=0.49\textwidth]{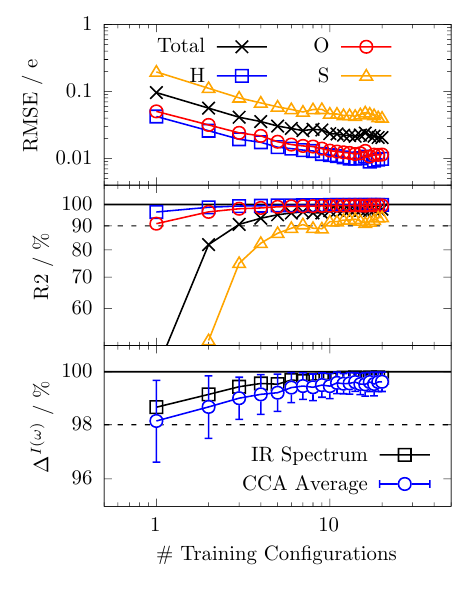}
    \caption{
        APTNN performance as a function of training set size for a \ce{SO4^{2-}} ion dissolved in liquid water, 
        quantified by the element-wise RMSE (top panel), 
        the corresponding $R^2$ value (\eref{eq:R2}, center panel), 
        and the predicted IR spectrum compared to the explicit \emph{ab initio} reference (bottom panel).
        Note the logarithmic x- and y-scale in all panels.
        The predicted IR spectra are compared both at the level of the total spectrum (black open rectangles) 
        and at the level of the CCA decomposition (\eref{eq:cca_ir_lineshape}), in which the spectrum is partitioned into contributions from the solute S and O atoms, first- and second-shell water molecules, and all water molecules beyond the second shell. 
        For each CCA contribution, a deviation score relative to the corresponding \emph{ab initio} reference is computed (\eref{eq:lineshape_difference}); 
        the reported value (blue open circles) is the average of these scores, 
        while the error bars represent the standard deviation across all contribution scores, indicating their spread.
    }
    \label{fig:aptnn_convergence_rmse_so4}
\end{figure}

We exemplify the convergence of the APTNN, particularly regarding the computation of the vibrational spectra, using \ce{SO4^{2-}(aq)}.
This system features a rare species (the \ce{SO4^{2-}} ion) and different solvent environments, such as the solvation shells around the ion and bulk water. 
The corresponding figures for bulk liquid water are presented in the SI.
We present the learning curves of the APTNN in terms of the RMSE and the $R^2$ coefficient in \fref{fig:aptnn_convergence_rmse_so4} top and center panels, respectively.
The RMSEs and $R^2$ values are computed for an independent test set, randomly sampled from the trajectories, consisting of 100~configurations.
Note the logarithmic scale and the typical almost linear decrease of the RMSE in the lower panel of \fref{fig:aptnn_convergence_rmse_so4} as expected.
We find the total RMSE decreasing from 0.1~($R^2=\SI{44}{\percent}$) to 0.02~e~($R^2=\SI{98}{\percent}$) using 10 to 200 training configurations, providing all APTs per configuration. 
Each configuration contains 197 atoms in total, 128xH, 68xO, and 1xS atoms. 
The 
$R^2$ of S atoms, being the least abundant species, performs worst, decreasing from 0.2 ($R^2 < 0$) to 0.04~e ($R^2 = \SI{94}{\percent}$).
In \fref{fig_si:scatter_apt_so4} in the SI we present the corresponding parity plots as a function of training set size, 
which visually illustrate the model performance and confirm the excellent agreement between predicted and reference values.
In panel (c) of the same figure we present the percentage spectral difference, \eref{eq:lineshape_difference}, between the total spectra obtained by DFT and by the APTNN model as a function of the training set size. 
This comparison is performed on \emph{one} MLMD trajectory along which APTs have been computed explicitly by DFT using electric field derivatives throughout and predicted by the APTNN at different training set sizes, respectively.
For 10~training configurations only, we find an overall $R^2$ value of \SI{44}{\percent} for the model accuracy with respect to the DFT test set.
In the context of the CCA (\sref{sec:cca}) we discussed that the spectral response by rare species, being the \ce{SO4^{2-}} ion in this case, 
can be dominated by the water spectrum due to their small prevalence.
We therefore apply the CCA to the IR spectrum, \eref{eq:cca_ir_lineshape}, to dissect the spectrum into individual contributions stemming from the ion itself, the first and second solvation shell waters, and water molecules beyond the second solvation shell, following the topological decomposition outlined in Ref.~\cite{Schienbein-2017-JPCL}.
All resulting auto and cross correlations are individually compared between the DFT reference spectrum and the predicted APTNN spectrum using \eref{eq:lineshape_difference}.
In panel (c) of \fref{fig:aptnn_convergence_rmse_so4} we show the average score~-- and its corresponding standard deviation indicated by the error bars~-- of all these individual differences, ranging from 98 to \SI{99.6}{\percent} using 10 to 200 training configurations, respectively.
Remarkably, for 10~training configurations only, the CCA score for the sulfate ion spectrum only, i.e.\ only calculating the auto correlation of the sulfate ion, yields an agreement of \SI{99.1}{\percent}, although the relative RMSE of the sulfur atom is below \SI{50}{\percent} and thus far from being accurate.

\begin{figure}
    \begin{subfigure}{0.49\textwidth}
        \includegraphics[width=\textwidth]{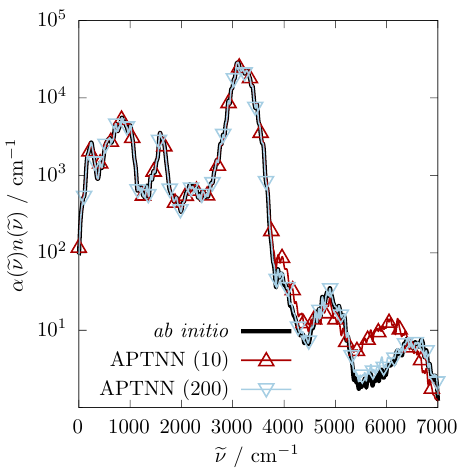}
        \caption{}
        \label{fig:aptnn_convergence_spectra}
    \end{subfigure}
    \hfill
    \begin{subfigure}{0.49\textwidth}
        \includegraphics[width=\textwidth]{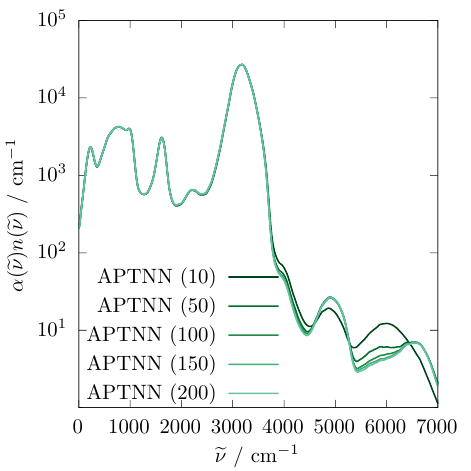}
        \caption{}
        \label{fig:aptnn_convergence_spectra_nodft}
    \end{subfigure}
    \caption{
        (a) Comparison of the explicit \emph{ab initio} spectrum (black) with machine-learning spectra obtained from APTNNs trained on 10 (red, upward triangles) and 200 training configurations (blue, downward triangles). 
        All spectra are computed from the same 20~ps MLMD trajectory, with APTs evaluated either from DFT reference calculations or predicted by the trained models.
        (b)
        Predicted spectra obtained from 80 independent 20 ps MLMD trajectories, using APTs evaluated from APTNNs trained on 10, 50, 100, 150, and 200 training configurations (light to dark green). 
        Note that the spectra are at the BLYP GGA functional level of theory and a Hann window of 0.5~ps has been applied.
    }
    \label{fig:aptnn_convergence_spectra_so4_h2o}
\end{figure}

While this spectral disagreement represents how well the spectrum is reproduced by the trained model on average, we translate that number into a frequency-dependent disagreement.
We present 
the total spectrum in \fref{fig:aptnn_convergence_spectra} obtained from the explicit DFT trajectory and the APTNN containing 10 and 200 configurations, respectively. 
Note the logarithmic scale of the absorption coefficient that is chosen since the predicted spectra are visibly indistinguishable from the DFT reference when using a linear scale. 
Even on the logarithmic scale, we find that the spectra look identical up to about \SI{4000}{\invcm}.
This frequency regime covers the most discussed and most intense features of liquid water, including the intermolecular H-bond stretch ($\approx\SI{200}{\invcm}$), the librational band ($\approx \SI{600}{\invcm}$), the intramolecular bending ($\approx\SI{1500}{\invcm}$), and the intramolecular OH stretch vibration ($\approx\SI{3500}{\invcm}$).
All features beyond \SI{4000}{\invcm} are overtones or combinations of these bands and therefore naturally come with orders of magnitudes less intensity.
Importantly, we find slight differences between the DFT reference and the APTNN (10~configurations) spectrum in the overtone at $\approx\SI{4900}{\invcm}$ and the overtone at $\approx\SI{6500}{\invcm}$ is not correctly captured. 
However, those overtones are much less studied in the literature since their intensities are several orders of magnitude smaller than the main features below \SI{4000}{\invcm}.
Remarkably, even those low intensity overtones are correctly reproduced when a training set containing 200~configurations is used.
In \fref{fig:aptnn_convergence_spectra_nodft} we present the final spectrum obtained from 80 times 20~ps MLMD trajectories, where APTs have been generated for all atoms along each trajectory using the trained APTNN.
The region below about \SI{4000}{\invcm} is already converged for a training set size as small as 10~configurations, while the overtones systematically converge as a function of training set size. 
At 50~configurations the frequency region up to \SI{5200}{\invcm} is correctly reproduced, while the region up to \SI{7000}{\invcm} is recovered using 150~training configurations. 

The iterative training procedure provides a practical early-stopping mechanism, as training can be halted once the predicted spectrum has converged within the frequency range of interest. 
Relying on spectral convergence alone can, however, be misleading: 
models that appear insufficiently accurate based on test-set metrics, 
for example, exhibiting total $R^2$ values as low as \SI{44}{\percent} with respect to an independent test set, 
may already reproduce IR spectra of the studied systems reasonably well. 
Conversely, a low test RMSE does not necessarily exclude more fundamental deficiencies in the model, such as non-locality or other systematic errors.
We therefore recommend a two-stage early-stopping strategy. 
In a first step, the model should be trained until test-set $R^2$ values on the order of \SI{90}{\percent} are reached for a sufficiently large independent validation set, as a coarse safeguard against major conceptual or numerical issues. 
In practice, such performance levels can typically be achieved with training sets comprising on the order of $10^2$ configurations, corresponding to roughly $10^3$ electronic-structure calculations~-- 
an effort that is generally tractable, particularly 
because all calculations can straightforwardly be parallelized.
In a second step, the iterative prediction of vibrational spectra can then be used to determine when further training 
does not alter the resulting spectrum any longer.

\subsection{Convergence of the PGTNN}

\begin{figure}
    \centering
    \includegraphics[width=0.49\textwidth]{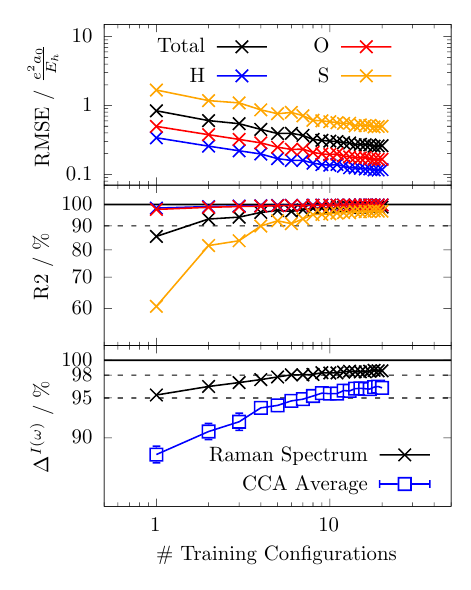}
    \caption{
        PGTNN performance as a function of training set size for a \ce{SO4^{2-}} ion dissolved in liquid water, 
        quantified by the element-wise RMSE (top panel), 
        the corresponding R2 value (\eref{eq:R2}, center panel), 
        and the score of predicted Raman spectra (\eref{eq:lineshape_difference}) compared to the explicit \emph{ab initio} reference (bottom panel).
        Note the logarithmic x- and y-scale in all panels.
        The predicted Raman spectra are compared both at the level of the total spectrum (black open rectangles) 
        and at the level of the CCA decomposition (\eref{eq:cca_raman_lineshape}), in which the spectrum is partitioned into contributions from the solute S and O atoms, first- and second-shell water molecules, and all water molecules beyond the second shell. 
        Furthermore we average the spectroscopic scores over the isotropic, parallel, and perpendicular scattering geometries.
        For each CCA contribution, a deviation score relative to the corresponding \emph{ab initio} reference is computed (\eref{eq:lineshape_difference}); 
        the reported value (blue open circles) is the average of these scores, 
        while the error bars represent the standard deviation across all contribution scores, indicating their spread.
    }
    \label{fig:pgtnn_convergence_rmse_so4}
\end{figure}

To demonstrate the convergence of the PGTNN, we repeat the same exercise as for the APTNN, for the aqueous \ce{SO4^{2-}} solution.
The corresponding figures for bulk liquid water are shown in the SI.
We present the learning curve of the PGTNN in terms of the RMSE and the $R^2$ coefficient in \fref{fig:pgtnn_convergence_rmse_so4} top and center, respectively.
Note the logarithmic scale in all figure panels.
The RMSEs and $R^2$ values are computed for an independent test set, randomly sampled from the trajectories, consisting of 100~configurations. 
The RMSE in the top panel decreases almost linearly as a function of training set size as expected.
We find the total RMSE decreasing from 
\num{0.84}~($R^2=\SI{85.5}{\percent}$) to 0.26~$e^2a_0/E_h$~($R^2=\SI{98.8}{\percent}$).
As for the APTNN, the S atom performs worst, decreasing from 
\num{1.68}~($R^2=\SI{60.6}{\percent}$) to 0.50~$e^2a_0/E_h$~($R^2=\SI{96.8}{\percent}$), 
because it is least abundant.
In \fref{fig_si:scatter_pgt_so4} in the SI we present the corresponding parity plots as a function of training set size, 
which visually illustrate the model performance and confirm the excellent agreement between predicted and reference values.
In panel (c) we present the percentage spectral agreement, \eref{eq:lineshape_difference}, between the total spectra obtained by DFT and by the PGTNN model as a function of the training set size.
We average over all Raman lineshape functions that depend on the different components of the polarizability tensor, see \eref{eq:raman_lineshape_cartesian}.
For 10 training configurations only, we find an average spectral agreement of \SI{95.4}{\percent} which increases to \SI{98.6}{\percent} when training on 200~configurations.
We employ the CCA again to decompose the Raman response, see \eref{eq:cca_raman_lineshape}, into topological groups, namely the \ce{SO4^{2-}} ion, the first and second solvation shells, and all remaining water molecules.
All resulting auto and cross correlations for all Raman lineshape functions are individually compared between the DFT reference and the predicted PGTNN spectra using \eref{eq:lineshape_difference} and averaged.
In panel (c) we show the average score, and its corresponding standard deviation indicated by the error bars~-- of all these individual percentages increasing from 88.0 to \SI{96.3}{\percent} for 10 and 200 training configurations, respectively.
The CCA score for the sulfate ion spectrum only, i.e.\ only calculating the auto correlation of the sulfate ion, yields an agreement of \SI{80.3}{\percent} and continuously improves as a function of training set size to \SI{96.9}{\percent} at 200 training configurations.

\begin{figure}
    \begin{subfigure}{0.49\textwidth}
        \includegraphics[width=\textwidth]{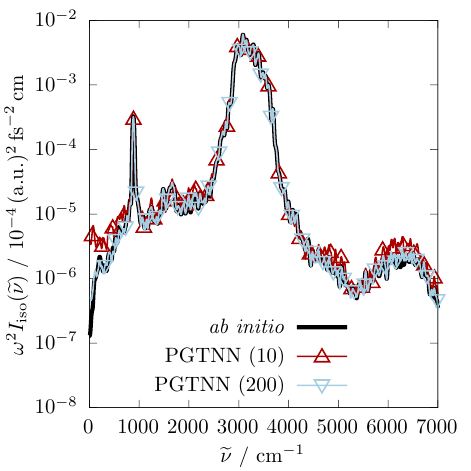}
        \caption{}
        \label{fig:pgtnn_convergence_spectra}
    \end{subfigure}
    \hfill
    \begin{subfigure}{0.49\textwidth}
        \includegraphics[width=\textwidth]{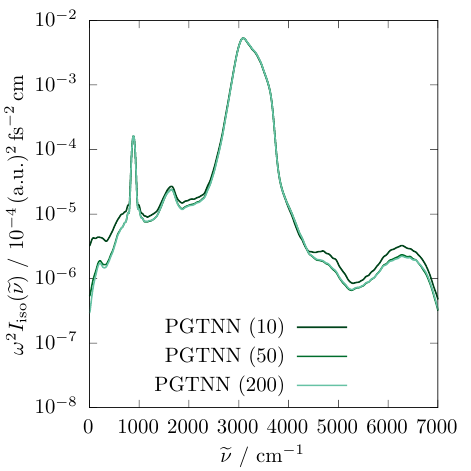}
        \caption{}
        \label{fig:pgtnn_convergence_spectra_nodft}
    \end{subfigure}
    \caption{
        (a) Comparison of the explicit \emph{ab initio} spectrum (black) with machine-learning spectra obtained from PGTNNs trained on 10 (red, upward triangles) and 200 training configurations (blue, downward triangles). 
        All spectra are computed from the same 20~ps MLMD trajectory, with PGTs evaluated either from DFT reference calculations or predicted by the trained models.
        (b)
        Predicted spectra obtained from 80 independent 20 ps MLMD trajectories, using PGTs evaluated from PGTNNs trained on 10, 50, 100, 150, and 200 training configurations (light to dark green). 
        Note that the spectra are at the BLYP GGA functional level of theory and a Hann window of 0.5~ps has been applied.
        Here ``(a.u.)'' refers to the atomic unit of electric polarizability ($e^2 a_0^2 / E_h$).
    }
    \label{fig:pgtnn_convergence_spectra_so4_h2o}
\end{figure}

To translate these condensed numbers into a frequency-dependent disagreement we show the isotropic Raman lineshape, \eref{eq:raman_lineshape_berne_iso}, in \fref{fig:pgtnn_convergence_spectra} obtained from the explicit DFT trajectory and the PGTNN containing 10 and 200 configurations, respectively.
Note the logarithmic scale of the figure.
We find that the PGTNN containing 10 training configurations only shows some disagreement over the full presented frequency range, particularly, where the intensity is on the order of $\num{1e-4}~\ramanint$ or less.
We find that the predicted isotropic Raman lineshape exactly agrees with the explicit DFT reference when using a PGTNN trained on 200~configurations. 
In \fref{fig:pgtnn_convergence_spectra_nodft} we present the final spectrum obtained from 80 times 20~ps MLMD trajectories, where PGTs have been generated for all atoms along each trajectory using the trained PGTNN.
We again find a systematic convergence of the spectrum as a function of training set size. 
Remarkably, the spectra are visually indistinguishable for the PGTNNs trained on 50 and 200 configurations.
The corresponding figures for the parallel~(VV) and perpendicular~(VH) Raman lineshapes are shown in the SI.

The presented data largely supports the same conclusion as for the APTNN elaborated above.
The iterative training procedure provides a practical early-stopping mechanism as well, where the training can be stopped once the predicted spectrum has converged within the frequency range of interest.
We again recommend that the model is trained until a test-set $R^2$ values on the order of \SI{90}{\percent} are reached to ensure that no systematic errors are present and the numerical accuracy/precision of the provided training data is sufficient.
In practice, such performance levels can typically be achieved with training sets comprising on the order of $10^2$ configurations as for the APTNN.
However, this time the electronic structure calculations are somewhat more expensive, since 13 electronic structure calculations are required per configuration.
The computational effort to generate the reference data is thus about a factor of 2 more expensive compared to the APTNN~-- an effort that is still very tractable and recall that some single point calculations can be shared between APT and PGT calculations, see \sref{sec:electronic_structure}.

\subsection{Spectral Assignment}

\begin{figure}
    \centering
    \includegraphics[width=0.49\textwidth]{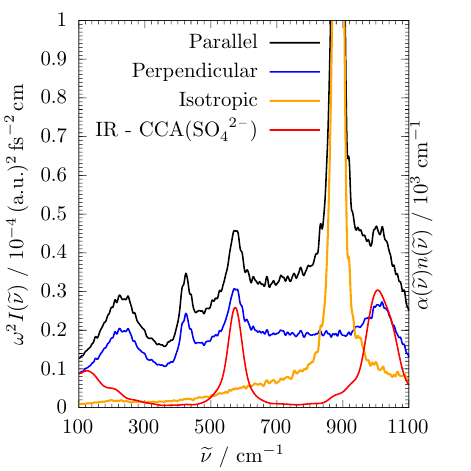}
    \caption{
        Isotropic (\eref{eq:raman_lineshape_berne_iso}, orange), parallel (\eref{eq:raman_lineshape_berne_vv}, black), and perpendicular (\eref{eq:raman_lineshape_berne_vh}, blue) Raman spectra of the aqueous \ce{SO4^{2-}} ion based on 80 NVE trajectories, 20~ps each, where the final PGTNN trained on 200 configurations has been employed. 
        For comparison the IR absorption coefficient of the sulfate ion only obtained from the CCA decomposition (\eref{eq:cca_ir_cf}, red, right y axis) is shown, where the final APTNN trained on 200~configurations has been employed as well.
        Here ``(a.u.)'' refers to the atomic unit of electric polarizability ($e^2 a_0^2 / E_h$).
    }
    \label{fig:raman-flavors}
\end{figure}

Having established that APTNNs and PGTNNs accurately reproduce IR and Raman spectra at the underlying DFT level, we next assess whether the decomposition of vibrational spectra into Raman scattering geometries and atomic contributions is physically consistent. 
The purpose of this analysis is not to perform a detailed spectroscopic assignment~--
that is generally possible in condensed-phase spectra, see e.g.\ Refs.~\cite{Schienbein-2020-ANIE, Joll-2025-JPCL-THz}~--
but rather to validate that the proposed methodology correctly resolves polarization-dependent Raman responses and yields meaningful atom-resolved spectral signals. 
This cannot be achieved by comparing ML spectra with explicit DFT spectra, since both are dissected consistently, such that an external reference is required. 
The aqueous sulfate solution is a prime example for that exercise since IR and Raman spectra have been measured and normal modes have been calculated in the past~\cite{Rudolph-1999-ZPhysChem, BenMabrouk-2013-JRamanSpec}.
\fref{fig:raman-flavors} shows the parallel, perpendicular, and isotropic Raman spectra, together with the 
sulfate-specific IR response that is obtained by employing the atomic CCA, \eref{eq:cca_ir_lineshape}, 
and taking the auto correlation of the sulfate ion only.
The isotropic Raman spectrum shows one single band at around \SI{900}{\invcm} that is also present in the parallel spectrum, but missing in the perpendicular and IR spectrum.
The parallel and perpendicular Raman spectra show three prominent bands at around \num{250}, \num{450}, and \SI{590}{\invcm}, that are all missing in the isotropic Raman spectrum. 
We further observe one signal at around \SI{1050}{\invcm} in the parallel and perpendicular Raman spectrum.
These characteristic features are fully consistent in shape, relative intensity, and their appearance different scattering geometries with experimentally reported Raman spectra~\cite{Rudolph-1999-ZPhysChem}.
This confirms that 
we can correctly recover
isotropic, anisotropic, parallel, and perpendicular Raman spectra utilizing the PGTNN.
The sulfate-specific IR response shows two features at \num{590} and \SI{1050}{\invcm}, while the other Raman features remain fully silent. 
This is consistent with selection rules determined from a normal mode analysis of the sulfate ion only~\cite{BenMabrouk-2013-JRamanSpec} 
and confirms that the CCA framework is a powerful tool to isolate spectral contributions of specific substances that are otherwise buried below a strong background.

\section{Conclusions}

In this work, we introduced \texttt{mimyria}, an automated and modular software framework that enables the generation, validation, and analysis of IR and scattering-geometry dependent Raman spectra from molecular dynamics trajectories.
The framework provides a complete end-to-end workflow, spanning the preparation of electronic-structure reference calculations (via ready-to-use templates for \texttt{CP2k}), automated training of machine-learning response models, spectral prediction, and quantitative validation. 
By integrating these components into a unified pipeline, \texttt{mimyria} lowers the practical barrier for routinely applying vibrational spectroscopy in condensed-phase simulations.
A central methodological contribution of this work is the introduction of the polarizability gradient tensor (PGT) as an atom-resolved response property for Raman spectroscopy, complementing the established atomic polar tensor (APT) for IR spectroscopy. 
We demonstrate that both APTs and PGTs can be computed consistently from electronic-structure theory, validated across formally equivalent derivative routes, and learned efficiently by machine-learning models. 
Beyond technical correctness, APTs and PGTs represent physically meaningful target properties: they provide a rigorous route to decompose vibrational spectra into atomic contributions, 
for instance by the cross correlation analysis or solute-solvent separation allowing for spectral assignment and for isolating species-specific bands from a dominating background.
Because these tensors directly connect atomic motion to IR intensities and Raman lineshapes, they allow arbitrary velocity-correlation observables to be promoted to physically correct spectral responses.

Our benchmarks further show that machine-learned vibrational spectra converge 
rapidly with training set size,
and that spectral accuracy can be assessed without requiring explicit \emph{ab initio} reference spectra. 
Iterative training thus provides a natural early-stopping criterion, and training-set requirements can be tailored to the intended scientific objective: reproducing dominant spectral features requires comparatively little data, whereas atom-resolved decompositions or very low-intensity bands demand larger training sets. 
These trends provide practical guidance for deploying machine-learning–based spectroscopy in production workflows.
Vibrational spectra in this framework are derived from molecular dynamics and therefore anharmonicities, finite-temperature effects, and explicit solvent interactions are inherently included, enabling direct and, in principle, quantitative comparison with experimental measurements. 
Residual frequency shifts observed in the present examples primarily reflect the underlying electronic-structure reference rather than limitations of the machine-learning or decomposition methodology, underscoring the transferability of the approach across theoretical levels.

Taken together, this work establishes a validated, automated, and data-efficient route to IR and Raman spectroscopy from molecular dynamics, while demonstrating the broader scientific value of atom-resolved response tensors. 
We expect \texttt{mimyria}, together with APT- and PGT-based learning, to enable routine, quantitatively reliable vibrational spectroscopy for complex molecular systems, opening new avenues for spectroscopic interpretation, model validation, and experiment–theory integration.

\vspace{1cm}

\noindent\textbf{Acknowledgement}

We thank Christoph Schran (Cambridge) for providing the \texttt{CP2k} input files for the aqueous \ce{SO4^{2-}} solution.
This project was supported by the \textit{Deutsche Forschungsgemeinschaft} (DFG, German Research Foundation) under Germany's Excellence Strategy~-- EXC 2033~-- 390677874~-- RESOLV.

\clearpage
\printbibliography

\end{document}

\title{Supporting Information for:\\Mimyria: Machine learned vibrational spectroscopy for aqueous systems made simple}

\author[1,2]{Philipp Schienbein\thanks{email: philipp.schienbein@ruhr-uni-bochum.de}}
\affil[1]{Lehrstuhl für Theoretische Chemie II, Ruhr-Universität Bochum, 44780 Bochum, Germany}
\affil[2]{Research Center Chemical Sciences and Sustainability, Research Alliance Ruhr, 44780 Bochum, Germany}
\date{ }

\maketitle

\section{SCF precision for PGTs}

\begin{figure}[h]
    \centering
    \includegraphics[width=0.49\textwidth]{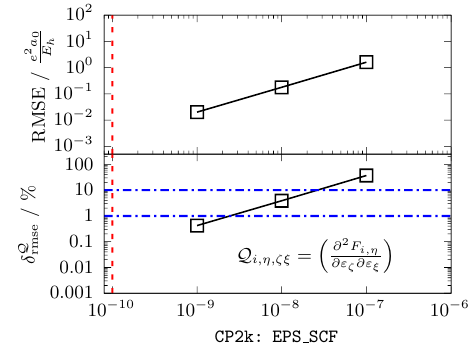}
    \caption{
        RMSE (top panel) and relative RMSE, 
        $\delta_\text{rmse}^\mathcal{Q}$, see \eref{eq:apt_rel_rmse} (bottom panel), 
        when computing PGTs from taking the derivative of atomic forces with respect to an electric field, \eref{eq:pgt-identity}, last term,
        as a function of the SCF convergence criterion in \texttt{CP2k}.
        An electric field strength of \num{5e-4} is used as displacement, see \fref{fig:pgt-efield-numderiv} in the main text.
%
        %or from taking the derivative of the polarizability tensor with respect to an atomic displacement in space (b), \eref{eq:pgt-identity}, central term,
        %both evaluated numerically by central finite differences using the displacement given at the respective abscissa.
        %
        %The electric field test is based on 30 configurations, while the spatial derivative test is based on a single configuration, all containing 128 water molecules each, see text.
        %
        The vertical red dashed line indicates the respective reference value, being \texttt{EPS\_SCF} equals \num{1e-10}.
        %s ($\approx \SI{0.026}{\VpA}$) in case of the electric field derivative and 
        %\SI{0.01}{\angstrom} in case of the spatial derivative.
        %
        The horizontal blue dashed-dotted lines in the bottom panel mark relative errors of \num{1} and \SI{10}{\percent}, respectively.
        %respective deviations when comparing the field with the spatial derivative, see \fref{fig:pgt-dft-rmse-comparison}.
    }
    \label{fig_si:pgt_eps}
\end{figure}

\clearpage

\section{Convergence of the APTNN: Aqueous Sulfate Solution}

\begin{figure}[h]
    \centering
    \includegraphics[width=0.49\textwidth]{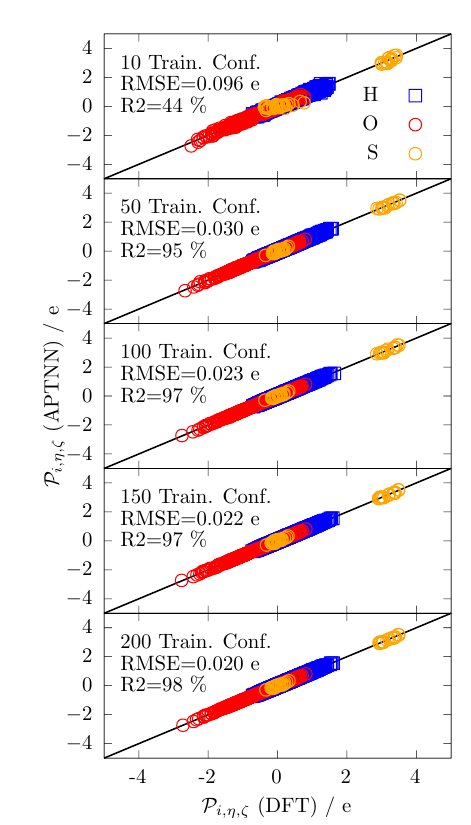}
    \caption{
        Parity plots quantifying the performance of the APTNN trained on pure liquid water as a function of the number of training configurations from 10~configurations (top) up to 200~configurations (bottom), 
        evaluated against an independent consistent test set containing 100~configurations.
        All components of the APT are shown.
        The figure complements the corresponding learning curve shown in \fref{fig:aptnn_convergence_rmse_so4} in the main text, where the RMSE and $R^2$ value are presented as a function of training set size. 
    }
    \label{fig_si:scatter_apt_so4}
\end{figure}

\clearpage

\section{Convergence of the APTNN: Pure Liquid Water}

\begin{figure}[h]
    \centering
    \includegraphics[width=0.49\textwidth]{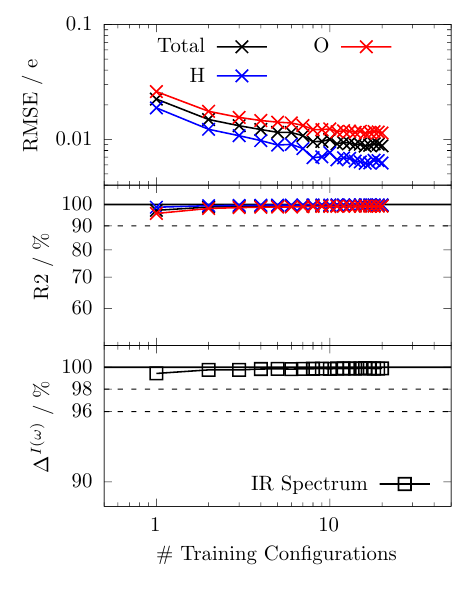}
    \caption{
        Same as \fref{fig:aptnn_convergence_rmse_so4} in the main text, but for bulk liquid water.
        %(a) and the aqueous \ce{F^-} solution.
        %
    }
    \label{fig_si:learning_curve_h2o}
\end{figure}

\begin{figure}[h]
    \begin{subfigure}{0.49\textwidth}
        \includegraphics[width=\textwidth]{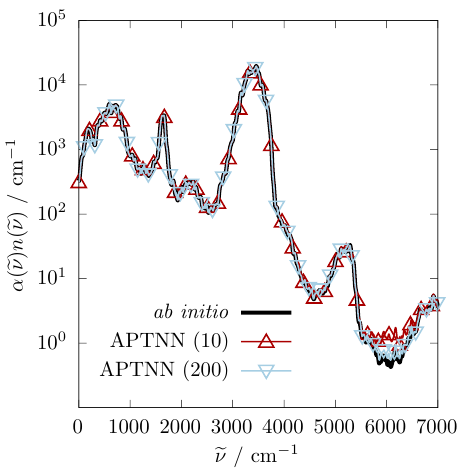}
        \caption{}
        \label{fig_si:aptnn_convergence_spectra_h2o}
    \end{subfigure}
    \hfill
    \begin{subfigure}{0.49\textwidth}
        \includegraphics[width=\textwidth]{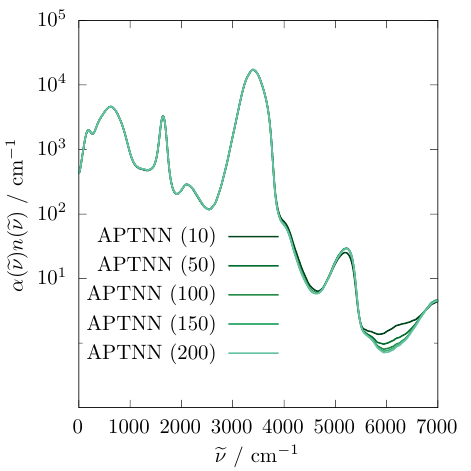}
        \caption{}
        \label{fig_si:aptnn_convergence_spectra_nodft_h2o}
    \end{subfigure}
    \caption{
        Same as \fref{fig:aptnn_convergence_spectra_so4_h2o} in the main text, but for bulk liquid water.
    }
\end{figure}

\begin{figure}[h]
    \centering
    \includegraphics[width=0.49\textwidth]{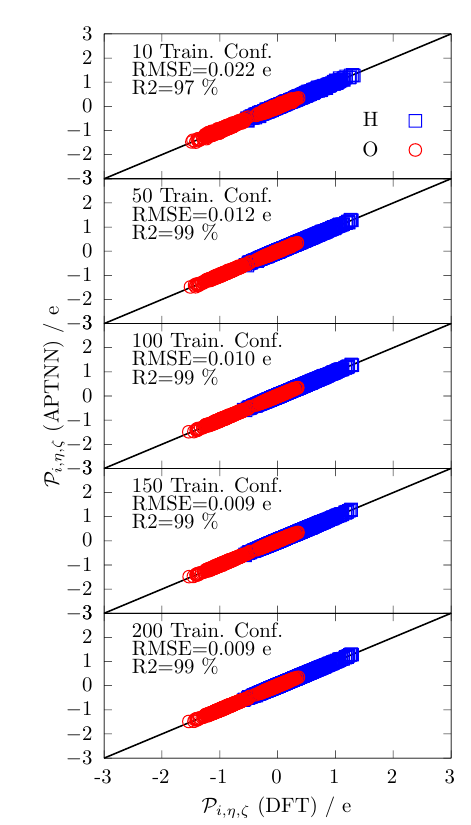}
    \caption{
        Parity plots quantifying the performance of the APTNN trained on pure liquid water as a function of the number of training configurations from 10~configurations (top) up to 200~configurations (bottom), 
        evaluated against an independent consistent test set containing 100~configurations.
        All components of the APT are shown.
        The figure complements the corresponding learning curve shown in \fref{fig_si:learning_curve_h2o}, where the RMSE and $R^2$ value are presented as a function of training set size. 
    }
    \label{fig_si:scatter_apt_h2o}
\end{figure}

\clearpage
\section{Convergence of the PGTNN: Aqueous Sulfate Solution}

\begin{figure}[h]
    \begin{subfigure}{0.49\textwidth}
        \includegraphics[width=\textwidth]{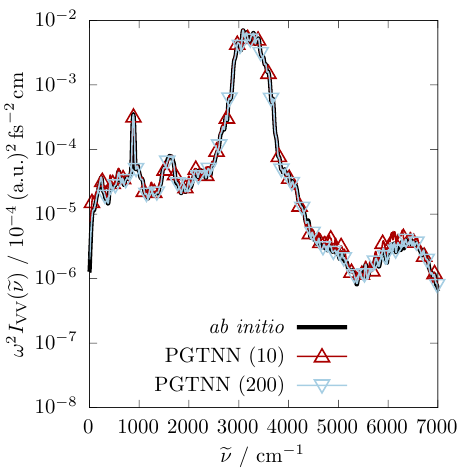}
        \caption{}
%        \label{fig_si:pgtnn_convergence_spectra_so4_vv}
    \end{subfigure}
    \hfill
    \begin{subfigure}{0.49\textwidth}
        \includegraphics[width=\textwidth]{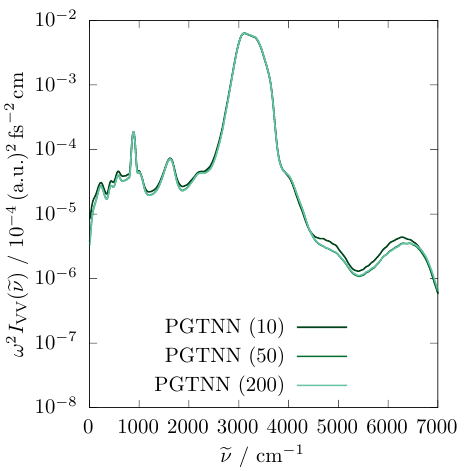}
        \caption{}
%        \label{fig:pgtnn_convergence_spectra_nodft}
    \end{subfigure}
    \caption{
        Same as \fref{fig:pgtnn_convergence_spectra_so4_h2o} in the main text, but for the parallel~(VV) Raman scattering geometry \eref{eq:raman_lineshape_berne_vv}.
    }
    \label{fig_si:pgtnn_convergence_spectra_so4_h2o_vv}
\end{figure}

\begin{figure}[h]
    \begin{subfigure}{0.49\textwidth}
        \includegraphics[width=\textwidth]{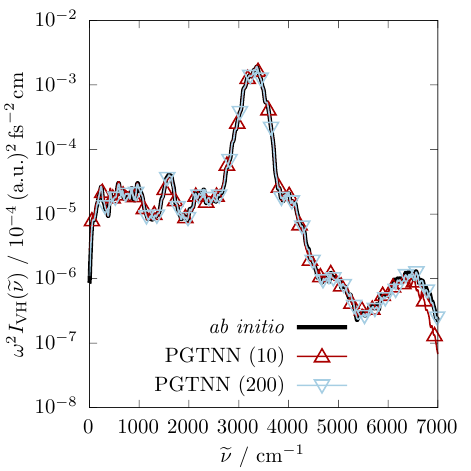}
        \caption{}
%        \label{fig_si:pgtnn_convergence_spectra_so4_vh}
    \end{subfigure}
    \hfill
    \begin{subfigure}{0.49\textwidth}
        \includegraphics[width=\textwidth]{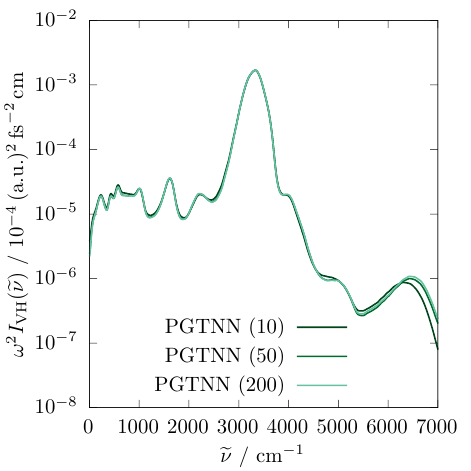}
        \caption{}
%        \label{fig:pgtnn_convergence_spectra_nodft}
    \end{subfigure}
    \caption{
        Same as \fref{fig:pgtnn_convergence_spectra_so4_h2o} in the main text, but for the perpendicular~(VH) Raman scattering geometry \eref{eq:raman_lineshape_berne_vh}.
    }
    \label{fig:pgtnn_convergence_spectra_so4_h2o_vh}
\end{figure}

\begin{figure}[h]
    \centering
    \includegraphics[width=0.49\textwidth]{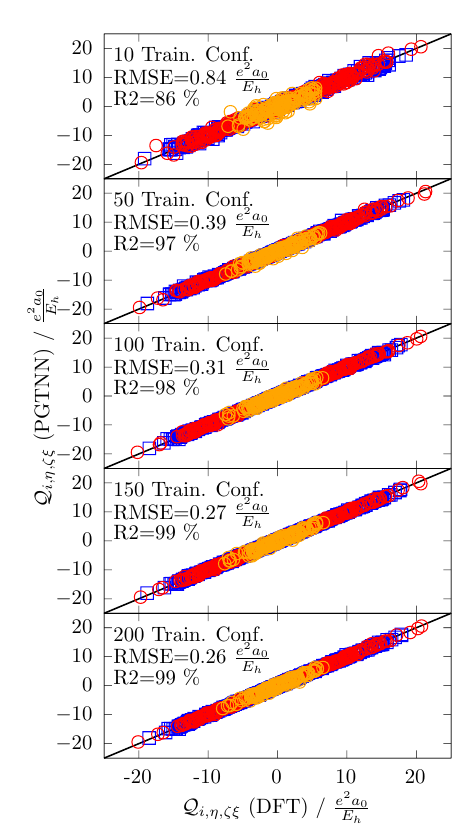}
    \caption{
        Parity plots quantifying the performance of the PGTNN trained on pure liquid water as a function of the number of training configurations from 10~configurations (top) up to 200~configurations (bottom), 
        evaluated against an independent consistent test set containing 100~configurations.
        All components of the PGT are shown.
        The figure complements the corresponding learning curve shown in \fref{fig:pgtnn_convergence_rmse_so4} in the main text, where the RMSE and $R^2$ value are presented as a function of training set size. 
    }
    \label{fig_si:scatter_pgt_so4}
\end{figure}

\clearpage
\section{Convergence of the PGTNN: Pure Liquid Water}

\begin{figure}[h]
    \centering
    \includegraphics[width=0.49\textwidth]{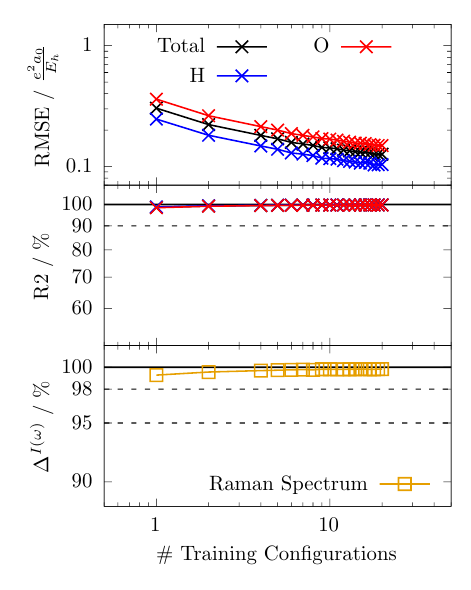}
    \caption{
        Same as \fref{fig:pgtnn_convergence_rmse_so4} in the main text, but for bulk liquid water.
    }
    \label{fig_si:learning_pgt_h2o}
\end{figure}

\begin{figure}[h]
    \begin{subfigure}{0.49\textwidth}
        \includegraphics[width=\textwidth]{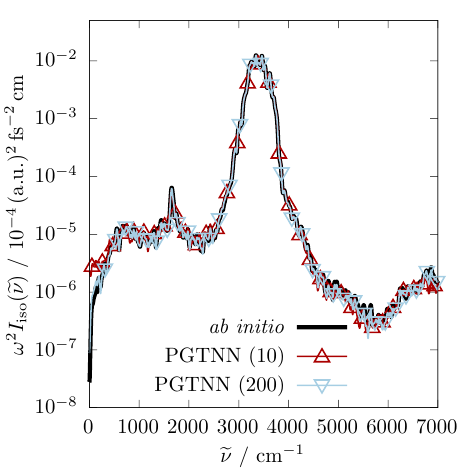}
        \caption{}
%        \label{fig_si:pgtnn_convergence_spectra_so4_vh}
    \end{subfigure}
    \hfill
    \begin{subfigure}{0.49\textwidth}
        \includegraphics[width=\textwidth]{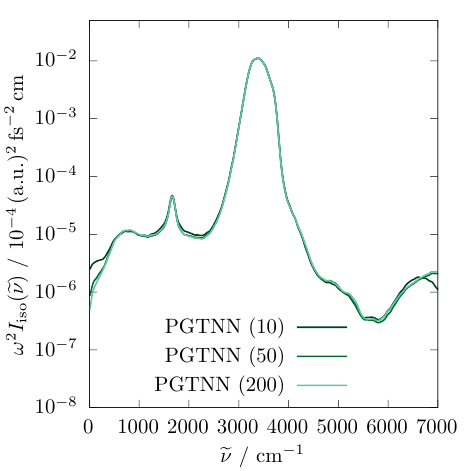}
        \caption{}
%        \label{fig:pgtnn_convergence_spectra_nodft}
    \end{subfigure}
    \caption{
        Same as \fref{fig:pgtnn_convergence_spectra_so4_h2o} in the main text, 
        %but for the perpendicular~(VH) Raman scattering geometry \eref{eq:raman_lineshape_berne_vh}.
        but for pure bulk liquid water.
    }
    \label{fig:pgtnn_convergence_spectra_h2o_iso}
\end{figure}

\begin{figure}[h]
    \begin{subfigure}{0.49\textwidth}
        \includegraphics[width=\textwidth]{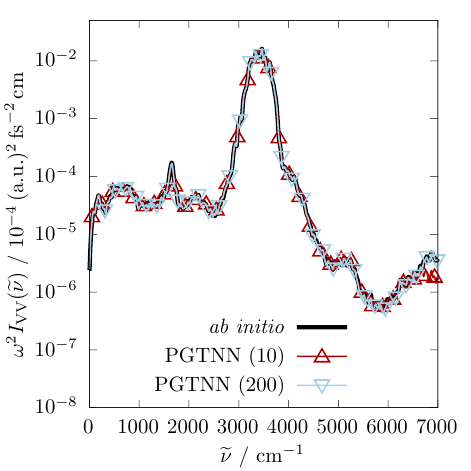}
        \caption{}
%        \label{fig_si:pgtnn_convergence_spectra_so4_vh}
    \end{subfigure}
    \hfill
    \begin{subfigure}{0.49\textwidth}
        \includegraphics[width=\textwidth]{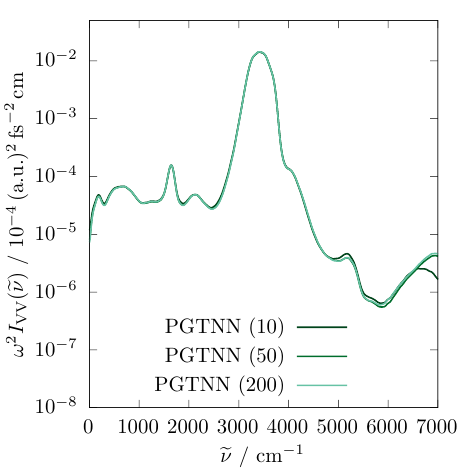}
        \caption{}
%        \label{fig:pgtnn_convergence_spectra_nodft}
    \end{subfigure}
    \caption{
        Same as \fref{fig:pgtnn_convergence_spectra_so4_h2o} in the main text, but 
        for pure bulk liquid water and 
        the parallel~(VV) Raman scattering geometry \eref{eq:raman_lineshape_berne_vv}.
    }
    \label{fig:pgtnn_convergence_spectra_h2o_vv}
\end{figure}

\begin{figure}[h]
    \begin{subfigure}{0.49\textwidth}
        \includegraphics[width=\textwidth]{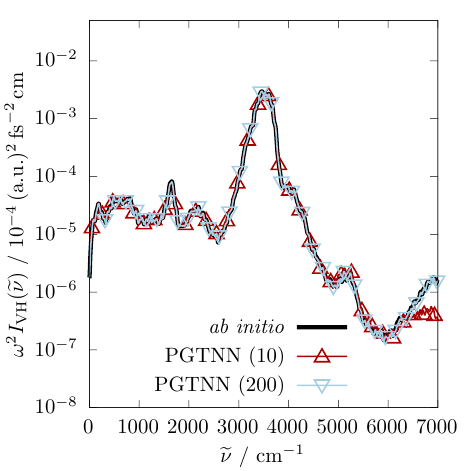}
        \caption{}
%        \label{fig_si:pgtnn_convergence_spectra_so4_vh}
    \end{subfigure}
    \hfill
    \begin{subfigure}{0.49\textwidth}
        \includegraphics[width=\textwidth]{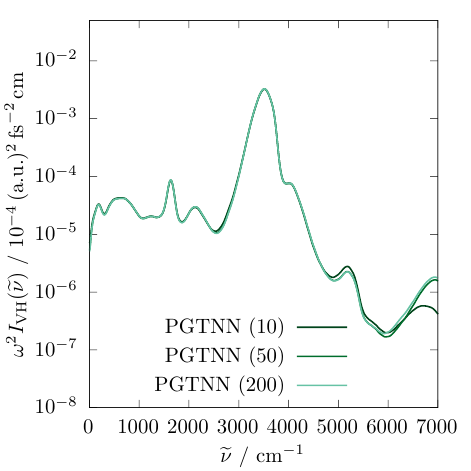}
        \caption{}
%        \label{fig:pgtnn_convergence_spectra_nodft}
    \end{subfigure}
    \caption{
        Same as \fref{fig:pgtnn_convergence_spectra_so4_h2o} in the main text, but 
        for pure bulk liquid water and 
        the perpendicular~(VH) Raman scattering geometry \eref{eq:raman_lineshape_berne_vh}.
    }
    \label{fig:pgtnn_convergence_spectra_h2o_vh}
\end{figure}

\begin{figure}[h]
    \centering
    \includegraphics[width=0.49\textwidth]{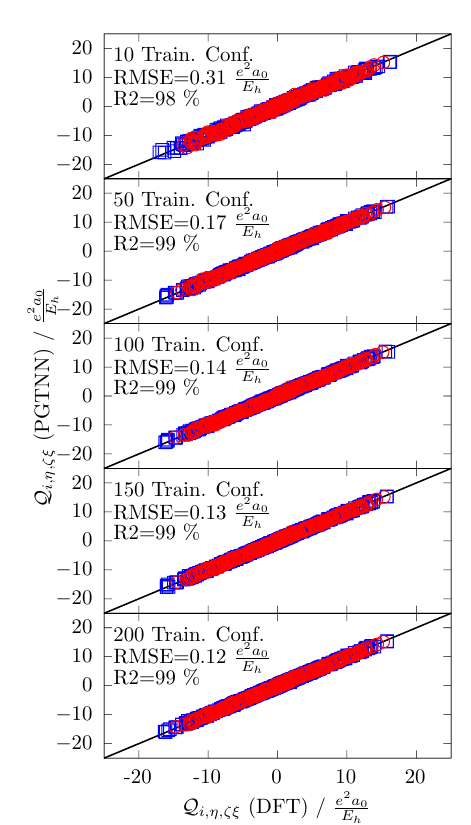}
    \caption{
        Parity plots quantifying the performance of the PGTNN trained on pure liquid water as a function of the number of training configurations from 10~configurations (top) up to 200~configurations (bottom), 
        evaluated against an independent consistent test set containing 100~configurations.
        All components of the PGT are shown.
        The figure complements the corresponding learning curve shown in \fref{fig_si:learning_pgt_h2o} in the main text, where the RMSE and $R^2$ value are presented as a function of training set size. 
    }
    \label{fig_si:scatter_pgt_h2o}
\end{figure}

\clearpage
\printbibliography